\documentclass[preprint]{aastex}

\shortauthors{Gizis, Reid, \& Hawley}
\shorttitle{M dwarfs and Local Star Formation}

\begin{document}

\title{The Palomar/MSU Nearby Star Spectroscopic Survey III: 
Chromospheric Activity, M-dwarf Ages and the Local Star Formation History
\footnote
{Observations were made at the 60-inch telescope at Palomar
Mountain which is jointly owned by the California Institute of Technology
and the Carnegie Institution of Washington}
}

\author{John E. Gizis}
\affil{Department of Physics \& Astronomy, University of Delaware
\email{gizis@udel.edu}}
\author{I. Neill Reid}
\affil{Space Telescope Science Institute, 3700 San Marin Drive, Baltimore, MD 21218 \\
and \\
Department of Physics and Astronomy, University of Pennsylvania,
209 South 33rd Street, Philadelphia PA 19104-6396\email{inr@stsci.edu}}
\author{Suzanne L. Hawley}
\affil{Department of Astronomy, Box 351580, University of Washington,
Seattle, WA 98195-1580\email{slh@astro.washington.edu}}

\begin{abstract}
We present high-resolution echelle spectroscopy of 676 nearby M dwarfs.
Our measurements include radial velocities, equivalent widths of
important chromospheric emission lines, and rotational velocities
for rapidly rotating stars.
We identify several distinct groups by their H$\alpha$ properties,
and investigate variations in
chromospheric activity amongst early (M0-M2.5) and mid (M3-M6) dwarfs.
Using a volume-limited sample together with a relationship 
between age and chromospheric activity,
we show that the rate of star formation in the immediate
Solar Neighbourhood has been relatively constant over the last 4 Gyr.
In particular our results are inconsistent with recent large
bursts of star formation.
We use the correlation between H$\alpha$ activity
and age as a function of colour to set constraints on the properties of
L and T dwarf secondary components in binary systems. 
We also identify a number of interesting stars, including rapid rotators,
radial velocity variables, and spectroscopic binaries.
\end{abstract}


\section{Introduction\label{intro}}

M dwarfs are the dominant stellar component of the Galaxy
by both number and mass. With main sequence lifetimes
much longer than the age of the universe, they are
a fair tracer of the overall properties of the Galactic disk.
Their chromospheric activity decays on
timescales of billions of years, providing 
an age indicator that is relevant for studies of Galactic evolution.
By relating the activity levels in M dwarfs
to age, we can measure the local star formation 
history.  The latter parameter is one of the major requirements in
modelling the local substellar mass function 
\citep{ldwarfmf}.  Moreover, M dwarf components in multiple systems
provide constraints on the age of the system. Finally,
an extensive sample of young, low-luminosity stars in the field can
furnish a prime hunting ground for imaging surveys designed to find
young, luminous brown dwarf and giant planet companions.

As with other low luminosity objects, detailed observations 
are only possible for M dwarfs in the immediate Solar Neighbourhood.
The most extensive source for such objects remains the preliminary version
of the third Nearby Star Catalogue  \citep[hereafter pCNS3]{gj91}. 
Papers I \citep{rhg95} and II \citep{hgr96} in this series describe
our moderate resolution ($\sim 3 $\AA), red ($6200-7500$\AA) spectroscopic
observations of candidate M dwarfs in the pCNS3. We have used
those spectra, together with data from the literature, to estimate
distances and spectral types. In Paper I we defined a volume-limited
sample of northern ($\delta > -30\arcdeg$) stars, and investigated the luminosity
function and kinematics of low-mass stars in the Galactic disk.
Paper II presented data for the  southern  stars and
investigated various aspects of the chromospheric behaviour of the whole sample using
the H$\alpha$ line as a diagnostic.\footnote{In both these papers,
the tables were printed incorrectly.  Combined tables are available
in electronic form at the ADC and CDS as \citet{pmsu97}.} 
In this paper, we present echelle observations of M dwarfs 
from the volume-limited sample defined in Paper I. 

Our echelle spectra are of high ($\sim .2 $\AA) resolution and cover
the wavelength range $\lambda\lambda 4800-9400$\AA\ for all stars, with
data extending to 3700\AA\ for a subset of the brighter stars. 
These observations encompass chromospheric emission lines due to hydrogen, helium, and 
ionised calcium. Previous 
surveys of activity in M dwarfs have generally concentrated on
earlier spectral types ($\le$ M4), as in the H$\alpha$ surveys of 202 stars 
by \citet{sh86}, or have been limited to relatively small samples,
such as the 24 late-type M dwarfs observed by \citet{gl86}. 
The only published analysis based on a volume-limited sample is
the recent work by \citet{dfpm98,dfbump99}, who present 
observations of field M dwarfs within eight parsecs.   
Their spectroscopy has higher resolution than our data, but their
sample includes only 118 stars.
Our survey therefore provides a more detailed
study than heretofore possible of the distribution of chromospheric activity  
amongst late-type dwarfs, and, combined with an age-activity relation calibrated
by M dwarfs in open clusters, the first opportunity to use those stars
to probe the Galactic disk star formation history.

The following section presents the basic analysis of our spectroscopic data, including
radial velocity and rotational velocity measurements and line strength determinations. 
Since we have multiple
observations of many stars, we have used our data to search for velocity
variations, and Section 3 identifies several previously-unrecognised 
binary systems.  Chromospheric activity is discussed in Section 4, 
where we investigate the H$\alpha$ properties of the sample, describe 
an age-activity relation calibrated with open cluster observations,
and use that relation to probe the recent
star formation history of the Galactic disk. Section 5 summarises our
main conclusions.

\section{Observations}

\subsection {The sample}

Our primary sample consists of the volume-complete sample of 499 single M dwarfs
and M dwarf primaries defined in Paper I. We refer to this as the VC sample.
The stars have absolute magnitudes
in the range $8 \le M_V \le 16$, declinations north of -30$^o$ and distances
within completeness limits ranging from 22 parsecs at M$_V = 8.5\pm0.5$ to 5 parsecs
at M$_V = 15.5\pm0.5$. The latter values were derived based on photometric and
trigonometric data available in late 1995, coupled with the (M$_V$, TiO5) relation
derived in paper I. Since then, Hipparcos-based trigonometric parallaxes 
\citet{hipparcos}, 
accurate to $\sim1$ milliarcsecond, have become available for approximately two-thirds of 
those systems. 
As will be discussed in Paper IV of this series (Reid {\sl et al.}, in preparation), 
the addition of the new astrometric data affects the inclusion/omission of 
only $\sim15\%$ of the stars in the Paper I sample.
While the new distance determinations should be taken into account in 
analysis of the kinematics of the local stars (Paper IV) 
they are of little importance for the
analysis of chromospheric activity and age that is undertaken in 
this paper. Thus, we retain the VC sample as our reference here. 

\subsection {Echelle spectroscopy}

Our data were obtained with the echelle spectrograph \citep{m85}
on the Palomar 60-inch telescope, which has a 2-pixel resolution of 19,000.
Observations between May 1994
and February 1995 used the original quartz $60\arcdeg$ cross-dispersing
prisms, giving wavelength coverage from $3700 $\AA\ to $9500 $\AA.
Beyond $7000 $\AA\ there are gaps in wavelength coverage where the orders
extend off the CCD.  Because our targets are all red stars, very few counts
were obtained in the blue ($\lambda < 4800$\AA) except for the very brightest
stars.  The result was that the useful part of the spectrum was squeezed
into the lower quarter of the CCD, leading to partial overlap between
the reddest adjacent orders.  To solve this problem,
beginning in June 1995, a new set of cross-dispersing
prisms (with SF3 glass and a $42\arcdeg$ apex) was used.  
This setup gives wavelength coverage from $4800$ to $9500 $\AA\
and has complete order separation, although the gaps in wavelength
coverage remain.  For both setups, the exposure time
was $600$ seconds per star, except for the faintest ($V>13$) 
stars for which
the exposure times were increased to as much as 1800 seconds. 

Data were extracted using the FIGARO echelle software 
(Tomaney \& McCarthy, private communication).
Wavelength calibration was determined from a 300-second
exposure of a Th-Ar lamp taken 
at the beginning of each night.  After each observation of 
a target star, a short (45 second) exposure of the arc lamp was taken 
at the same telescope position in order to remove instrumental
flexure.  The spectra were not flux calibrated.  

\subsection {Radial Velocities}

Radial velocities were determined by cross-correlation against  
reference M dwarfs from Marcy \& Benitz (1989 - MB89). The latter
velocities are accurate to better than 0.23 kms$^{-1}$, a factor of five
higher than the accuracy of our own observations (as discussed further below). 
For the bright, early M dwarfs (TiO5$>0.5$) we used the standard
echelle FIGARO cross-correlation program.  Each order
was correlated with the velocity standards, and the average
radial velocity from all the orders 
was determined.  The arc lamp exposures, taken adjacent to each
program star observation, 
were also cross-correlated, providing a correction for flexure.  

\begin{figure}
\epsscale{0.8}
\plotone{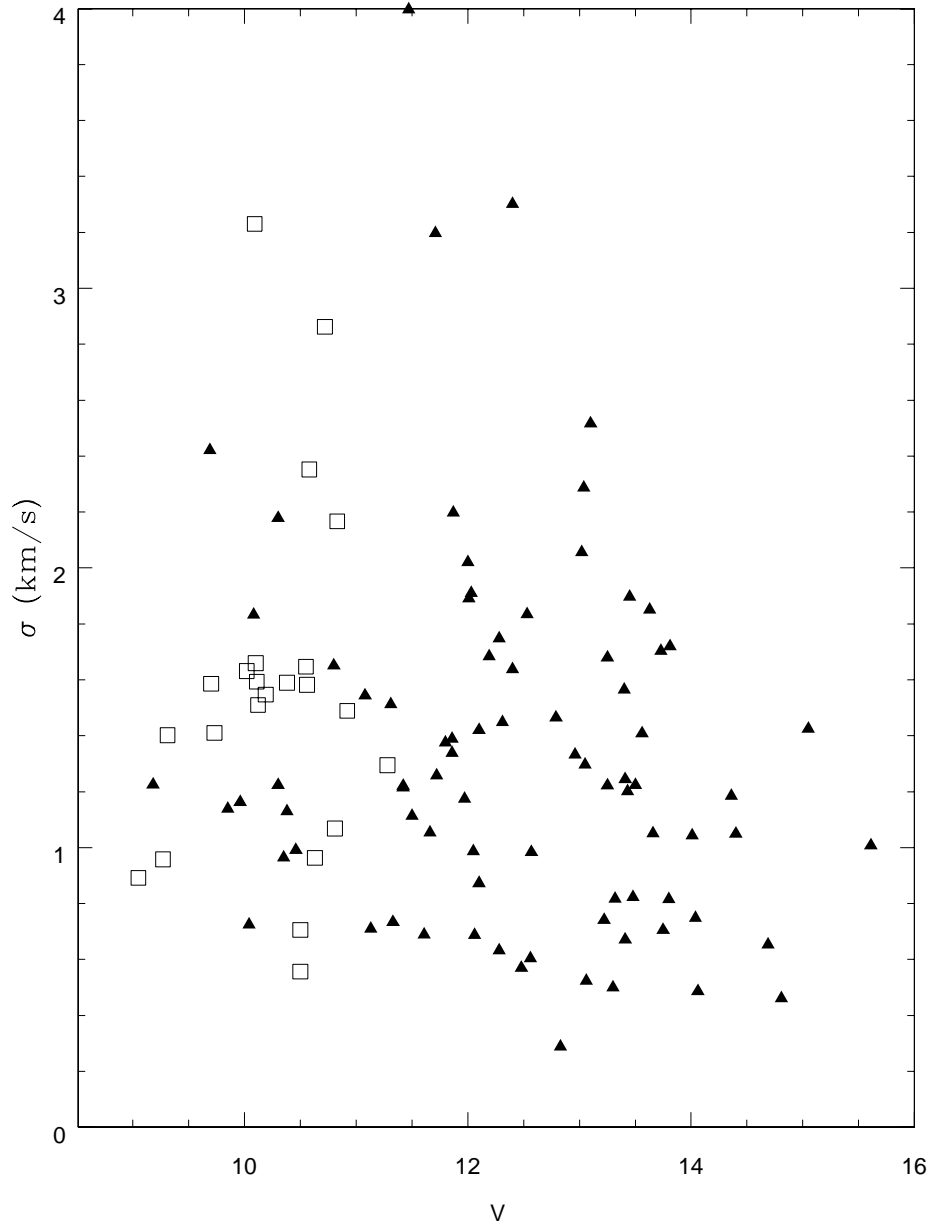}
\caption{The observed $\sigma$ of the measured radial velocities
for stars with multiple observations.  Early M dwarfs   
(${\rm TiO5}>0.5$) are shown as open squares and mid M dwarfs 
(${\rm TiO5} \le 0.5$) are shown as solid triangles.  There
is no trend with the apparent magnitude of the star.
\label{fig-sigmav}}
\end{figure}

This procedure did not provide reliable velocities 
for faint, late-type M dwarfs. The lower signal-to-noise at
blue wavelengths led to a higher potential for bias from the
effects of telluric absorption and night sky lines.
We were able to obtain reasonable results for those stars
by individually computing the cross-correlation for
each order and combining those measurements to find the median radial velocity.
This procedure was used for all stars with TiO5$\le 0.5$.  
Figure~\ref{fig-sigmav} plots the rms dispersion about the 
mean radial velocity for all M dwarfs with at least four measurements.
The distribution suggests a typical internal accuracy of 
$\lesssim$ 1.5 kms$^{-1}$. 

We can determine our external errors by comparison with previous
high accuracy velocity studies of M dwarfs.  The
results are shown in Table~\ref{table-vcomp}, where $\sigma_{ref}$
is the formal uncertainty of the reference sample. We note that
a comparison between \citet{toko88,toko92} and MB89
(13 stars) gives an rms of 
only $0.46$ kms$^{-1}$ but a mean difference of 0.78 kms$^{-1}$, consistent
with the offset derived from our observations, which are tied
to the MB89 system. Similarly, 
\citet{sh86} adopt a velocity of 14.1 kms$^{-1}$ for the velocity 
standard Gl 526, while MB89 measure a velocity of 15.7 kms$^{-1}$ for this
star. Again, the offset is consistent with our measurement. Finally, 
we list two comparisons with the recent observations by
Delfosse {\sl et al.} (1998) since the dominant contribution to the
residuals comes from three stars: Gl 206 (V$_{Del} = 8.0$kms$^{-1}$; $\Delta=V_{P60}-V_{Del}=9.2$
kms$^{-1}$), G165-008 (8.0; $\Delta=-15.5$) and Gl 268.3B (-6.0; $\Delta=-8.4$).
All three stars are known binaries, while G165-008 is also
a very rapid rotator (see further below). In general, the
comparison indicates that the velocities derived from our echelle spectra
are accurate to $\sigma < 1.5$ kms$^{-1}$.  

We can also examine the quality of the radial velocities found from
our previous moderate resolution spectra. Based on an
external comparison with the MB89 standards, we estimated an
accuracy of $\sim 15$ kms$^{-1}$ (Paper I). 
After excluding double-lined spectroscopic binaries,
velocity standards, and three stars with anomolously
large residuals, we have 582 stars in common.
The standard deviation is 17 kms$^{-1}$, confirming our previous estimate. 

The individual velocities and heliocentric Julian data for each observation, together
with the mean value and the rms dispersion based on our observations, 
are given in Table~\ref{table-velocities}. In addition to the star name, we list
the `NN' number cited in Papers I and II. While the latter is merely the rank order in
pCNS3, and is not an officially
recognised designation, it provides a straightforward means of cross-referencing
results between tables for this relatively large dataset. 
We also include measurements of three
M dwarfs which are not included in either the pCNS3 or Papers I and II, but 
which are discussed in  \citet{gr97}. Both the primary G 134-035 
and the secondary LP 197-12 were on the slit, so the reported measurement
is for a blend of the two stars.  G 173-018 is a short-period, 
double-lined dMe system.\footnote{Tables 2, 4 and 5, together with the 
data tables from Paper I and Paper II are available from our PMSU website
at: http://www.physics.upenn.edu/$~$inr/pmsu.html}. 

\subsection{Line Strengths\label{lines}}

\begin{figure}
\epsscale{0.8}
\plotone{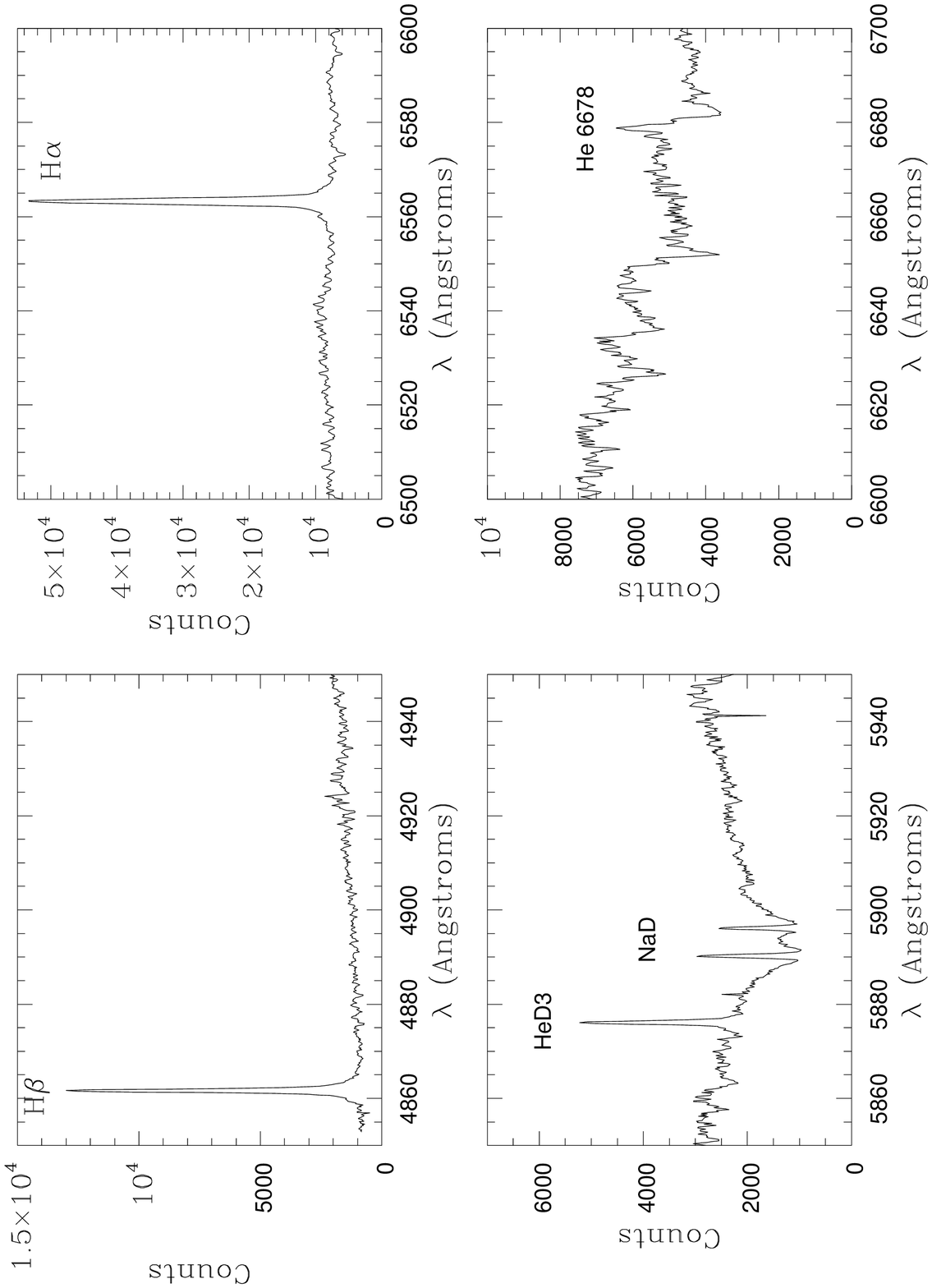}
\caption{Spectra of Gl 285 (NN 1219) showing the 
H$\alpha$, H$\beta$, He D3, NaD, and He 6678 emission lines.  
\label{fig-emlines}}
\end{figure}

To investigate chromospheric activity, we measured
the equivalent widths of atomic features using an 
automated program that counted the flux in rectangular passbands.   
The adopted
line (from F1 to F2) and pseudo-continuum (PC1 to PC2 and PC3 to PC4) 
regions are given in Angstroms in Table~\ref{table-defs},
and the average values observed for each star are given in
Table~\ref{table-widths}.
The appearance of the emission lines in a strong dMe star 
are illustrated in Figure~\ref{fig-emlines}.  For those sufficiently
bright stars observed with the old echelle prisms, the bluer Balmer
emission lines were also measured. 
The NaD lines were not measured
due to contamination by sky emission and the difficulty in defining
a pseudo-continuum.  The measurements of the He 6678 line were
difficult due to the weakness of the feature and the nearby TiO
absorption band.
We found that none of the stars had detectable lithium
absorption, as expected for M dwarf stars with ages 
exceeding $\sim 50$ Myrs.  

\begin{figure}
\epsscale{0.8}
\plotone{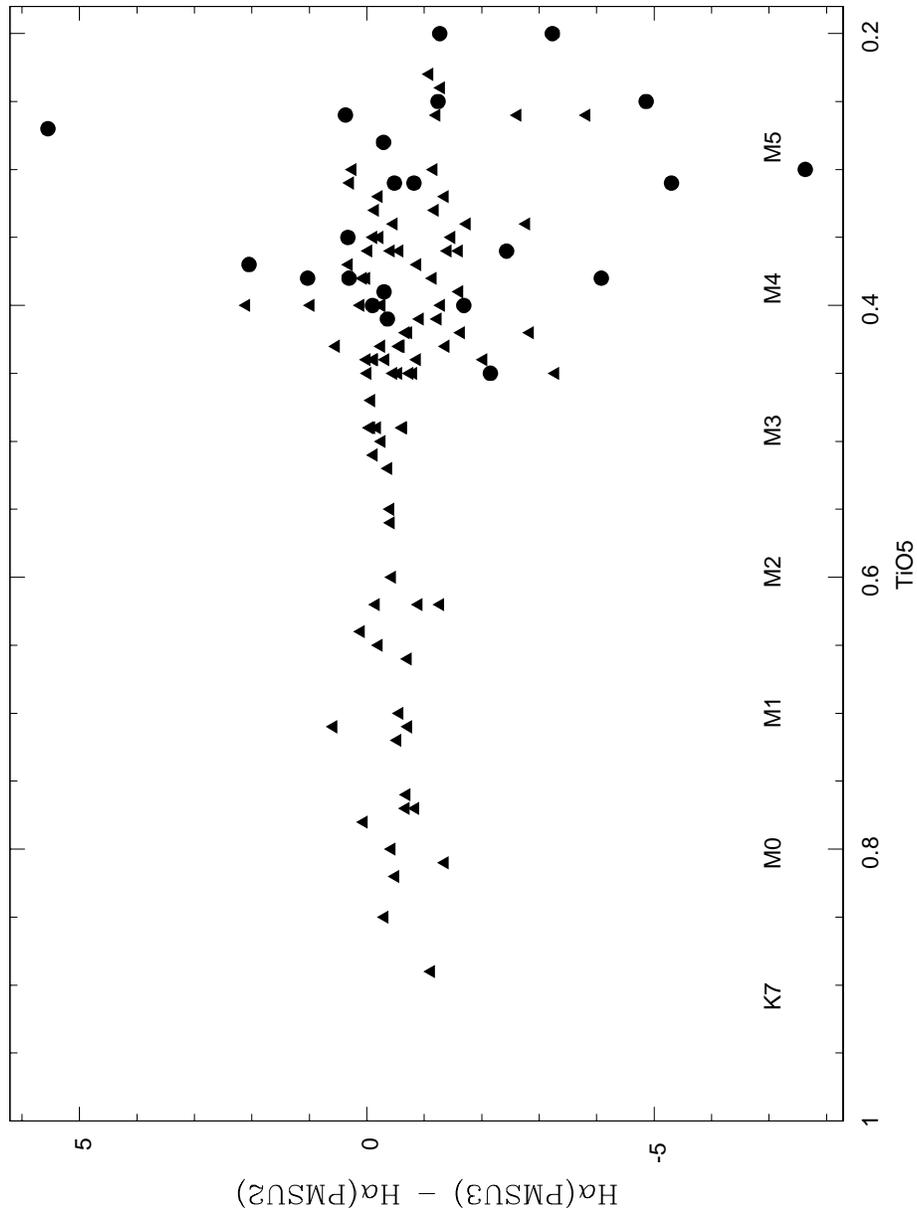}
\caption{A comparison between the H$\alpha$ emission line strength
measured at high resolution in this paper (PMSU3) and in our
previous, low-resolution work (Paper II).  Points where the PMSU3
measurements exceed 5\AA\ are shown as circles; weaker
dMe stars are shown as triangles.  A systematic difference
of $\sim 0.5$\AA\ is evident, which we attribute to differences
in placing the pseudo-continuum.  The increased scatter amongst later-type
stars is probably due to variability (see Section~\ref{activity} and
Figure~\ref{fig-hasig}).  
\label{fig-compha}}
\end{figure}

Figure~\ref{fig-compha} compares the H$\alpha$ equivalent width measured from
the echelle observations
with the lower resolution measurements from Paper I.  For the early
M dwarfs, the higher resolution observations are systematically
smaller by $\sim 0.5$\AA, probably reflecting differences in the placement of
the pseudo-continuum.  The large scatter for the cooler
dwarfs is due to the variability of these stars's chromospheres.  

\begin{figure}
\epsscale{0.8}
\plotone{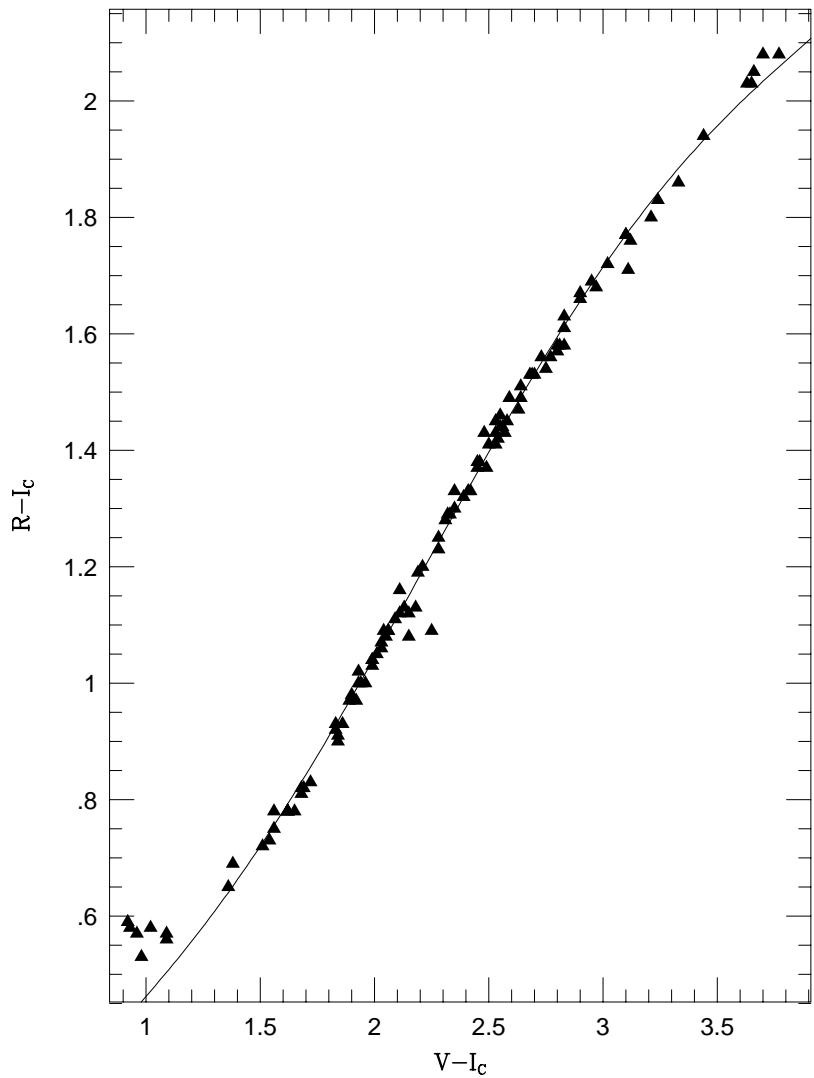}
\caption{Synthetic colors for M dwarfs using our low-resolution
spectra and the filter curves of \citet{b90b}.  The solid curve
is the \citet{b90a} polynomial fit to actual M dwarf photometry.  
We conclude that our spectra are correctly flux calibrated. 
\label{fig-color}}
\end{figure}

For most purposes, it is more useful to compare line fluxes rather
than equivalent widths. Absolute calibration is not available for all of the
stars in our sample, but the continuum flux near H$\alpha$ can be derived 
from broadband R$_C$ photometry \citep{rhm95}. To improve the
calibration, and extend it to H$\beta$, we 
obtained low resolution spectrophotometry of a subset
of 105 stars using the McCarthy spectrograph with a 150-line grating. 
The observations were made under photometric conditions
at the Palomar 60-inch on 9 and 10 July 1996.
The wavelength coverage was 4600\AA~to 10170\AA\ with
resolution  6\AA, providing a good match to the echelle observations.
In Figure~\ref{fig-color}, we show that use of the \citet{b90b} filter 
response curves
with our spectra reproduces the sixth order polynomial
relation between  $R-I_C$ and $V-I_C$ derived by \citet{b90a}.
The deviations seen at the blue end correspond to stars earlier than type
M0 and therefore do not effect our results.  

Using these low-resolution spectra, we find that the continuum flux at
H$\alpha$ and H$\beta$ can be derived from broadband photometry using the
following linear relations:
$$-2.5 \log F_{H\alpha} = (21.68 \pm 0.04) + (0.974 \pm 0.038) R_C$$
$$-2.5 \log F_{H\beta} = (21.01 \pm 0.06) + (1.055 \pm 0.055) V_C$$
where the fluxes are measured in erg s$^{-1}$ cm$^{-2}$ \AA$^{-1}$.  
These relations are applicable for stars with types M0-M6.5.  They are
used to transform the measured emission-line equivalent widths 
to the flux values discussed in Section~\ref{activity}.
We note that differences in continuum placement at low and high resolution
might lead to small systematic errors. 

V-band photometry is available for all the M dwarfs in our sample. In cases
where $R_C$ photometry was unavailable, we estimated the
$V-R_C$ color using the TiO5 value:
$$V-R_C = 3.2345 - 8.9529 \times {\rm TiO5} + 12.041 \times  {\rm TiO5}^2 - 5.6274\times {\rm TiO5}^3$$
When $V-I_C$ colors were needed but unavailable, we used:
$$V-I_C = 6.3657 - 14.6706 \times{\rm TiO5} + 17.6957 \times{\rm TiO5}^2 - 8.0934 \times {\rm TiO5}^3$$
Again these relations are applicable for stars with spectral types M0-M6.5.

\subsection{Rapid Rotators}

Rotational broadening is measurable for a handful of stars in our spectra.  
For each observing run, we artificially broadened a star of known low rotation 
($v \sin i < 2$ kms$^{-1}$) using
a rotational broadening function \citep{gray}.  We adopted
a limb darkening coefficient of 0.6, shown to be an appropriate value
by \citet{mc92}.  However, measuring  $v \sin i$ by cross-correlation against 
the standards proved ineffective, probably due to 
small mismatches in spectral type.  Instead,
we emulated the method of \citet{mc92} by using $\chi^2$
fitting to determine the best fit to individual atomic lines.
Our lower resolution requires that we use relatively strong atomic lines,
as in the region $6420 - 6520 $\AA.  Detections are limited to
$v \sin i > 20$ kms$^{-1}$ ($\sim 3$ pixels). We
estimate a measurement accuracy of $\sim 5$ kms$^{-1}$.

Stars with detected rotational broadening are listed in 
Table~\ref{table-rotation}.
Our results are quite repeatable; for example, G188-38 (NN 3453) was
observed 6 times, each time yielding $v \sin i$ between 35 and 40
kms$^{-1}$.  The most notable star is G 165-8 (NN 2128),
which has thirteen measurements that are all 
consistent with $v \sin i = 80$ kms$^{-1}$, 
similar to the extreme rotator Gl 890 \citep{petter}.
One of our observations clearly shows H$\alpha$ emission from a secondary companion, while
others show asymmetries in the line profile.  The rotational velocity is
close to that of the fastest-rotating M dwarfs in the Pleiades \citep{plvsini}, 
suggesting that the system is relatively young, $\tau \lesssim 10^8$ years.

A caveat to all of our rotation measures is that 
close binary pairs which happened to have velocity differences of $\sim 20$ kms$^{-1}$
at the time of our observations could be misidentified as rapid rotators. For
example, we measure $v \sin i = 30$ kms$^{-1}$ for Gl 735 (NN 2976), but 
we exclude the star from Table~\ref{table-rotation} since it is identified as ``SB''
in the pCNS3. Gl 206 (NN 933) and Gl 829 (NN 3360), both known SB2s, also 
show significant line broadening in our spectra.
Confusion is less likely where multiple measurements are available.

\section{Binaries} 

Table~\ref{table-sb2} lists 23 dwarfs which are
double-lined spectroscopic binaries (SB2).  These systems must have
relatively short periods, but our data provide insufficient
temporal coverage to attempt period determinations.   
Note that most objects in the present sample were
observed only once, so additional double-lined systems which
were in an unfavorable configuration at the time of our observation
remain undetected.
Thirteen of these 23 SB2 systems have published orbits, while the 
other ten are new discoveries. 

A few single lined stars with multiple velocity measurements, both
our own and those available in the literature, have velocities
spanning a larger range than expected given the individual uncertainties. 
This may reflect the presence of an unseen companion (i.e. SB1 system).  
Table~\ref{table-sb1}
lists stars where the overall rms dispersion of the observations exceeds 4 kms$^{-1}$.
Other stars listed in Table~\ref{table-velocities} with $\sigma > 3$ kms$^{-1}$ may also
merit additional observations.
Table~\ref{table-sb1diff} lists other candidate SB1 systems where a significant
velocity difference exists between our observation and the literature data.
With potential velocity amplitudes of a few kms$^{-1}$ and periods of
a few years, these systems are likely to have separations
of $\sim 1$ AU, equivalent to $\sim$ 0.1 arcseconds for stars in
this sample. These are therefore good candidates for high-resolution imaging 
(speckle, adaptive optics, or interferometry) and radial velocity monitoring
at higher resolution. Several stars have known companions which may be responsible for
the observed velocity differences.

\citet{poveda} have published a catalog of candidate wide binary or multiple
systems in the pCNS3 catalog. We have radial velocity measurements for
several of their candidates, including two with separations exceeding 0.1
parsec and therefore of potential interest for Galactic dynamics.  At
such separations, orbital velocities are less than 1 kms$^{-1}$, so we expect
the radial velocities of the components to agree within our measuring uncertainties. 
Table~\ref{table-wide} shows the comparison, where the final column gives the
observed velocity difference in terms of our measuring uncertainty, which 
we take as 1.5 kms$^{-1}$. Of the two wide systems, Gl 469 and Gl 471 are
clearly not associated (unless one has an unseen companion), but our data show
only a 3$\sigma$ difference between Gl 48 and the Gl 22 system. Except for
Gl 140 A/C, where orbital motion might affect the velocity of the brighter
star, the remaining candidate binaries have velocity differences consistent
with our measuring uncertainties. 

\section{Chromospheric Activity, Age and the Star Formation History}

The chromospheric age-activity correlation amongst main-sequence stars 
has been studied extensively for solar-type dwarfs, where it is
usually parameterised in terms of the `t to the half' law \citep{sk72},
\begin{displaymath}
F(Ca) \ = \ A t^{-1/2}
\end{displaymath}
where $F(Ca)$ is the Ca II K emission-line flux, $A$ is a constant and $t$ the age. 
Based on that calibration and observations of local G dwarfs, several groups have
attempted to reconstruct the recent star formation history in the disk.  In particular,
both \citet{barry} and \citet{rp00} claim that the data indicate several 
significant bursts of star formation over the last 4 Gyrs. On the other hand, 
\citet{sdj91} argue that these `bursts' are the result of a more complicated activity/age
relation than that implied by a simple Skumanich type power law.
Our goal is to use our observations of M dwarfs in the VC sample 
to address the local star formation history. To that end, we first
investigate the range of activity and the correlation of activity 
with age for the low-mass M dwarfs.  In particular, we do not
require that the M dwarfs follow the same age-activity relation
used for the G dwarfs.

\subsection{Activity\label{activity}}

\begin{figure}
\epsscale{0.8}
\plotone{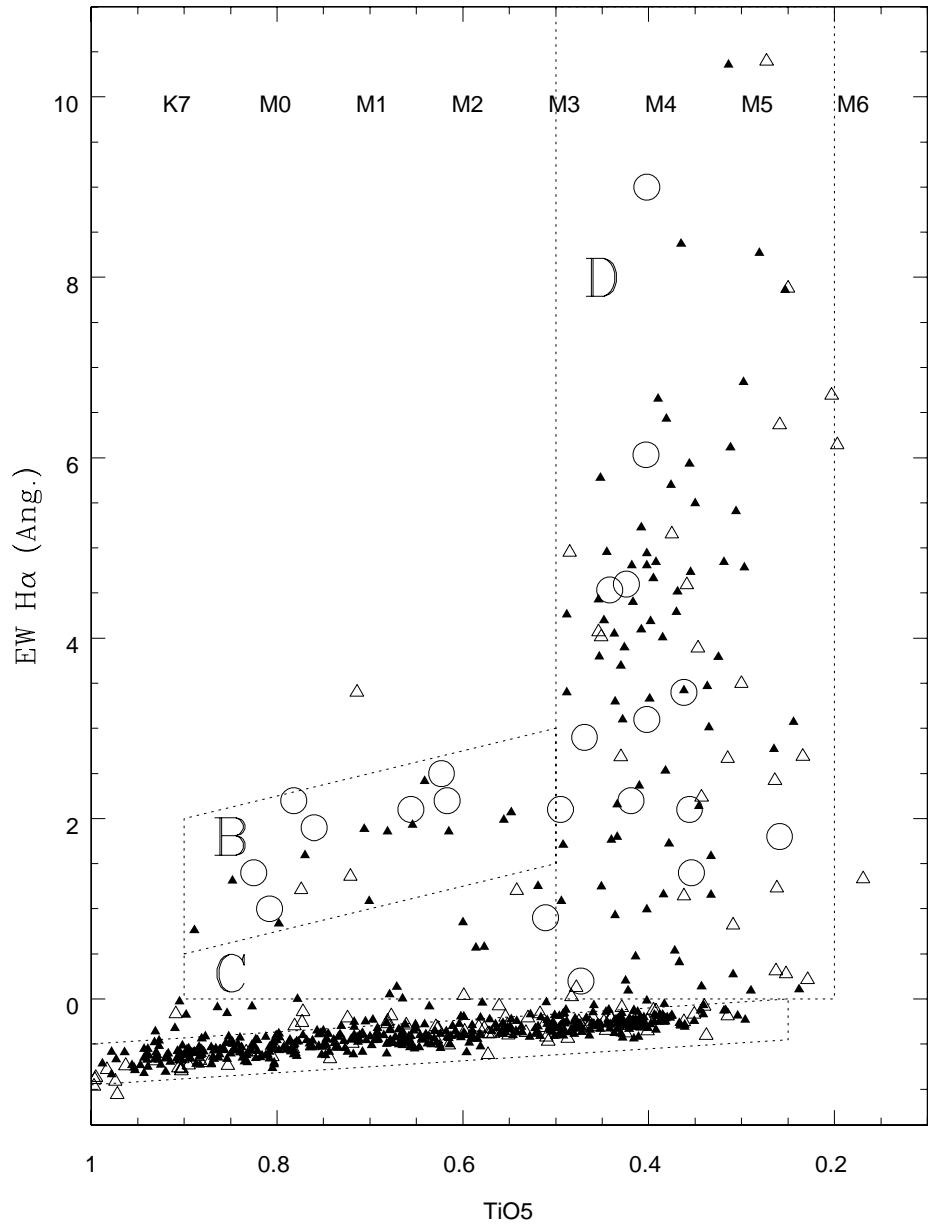}
\caption{The H$\alpha$ equivalent width as a function of TiO5 (spectral type).
Stars in the VC sample are shown as solid triangles, stars with emission
due to a close companion are shown as open circles, and other
non-VC stars are shown as open triangles.  \label{fig-tio5ha1}}
\end{figure}

\begin{figure}
\epsscale{0.8}
\plotone{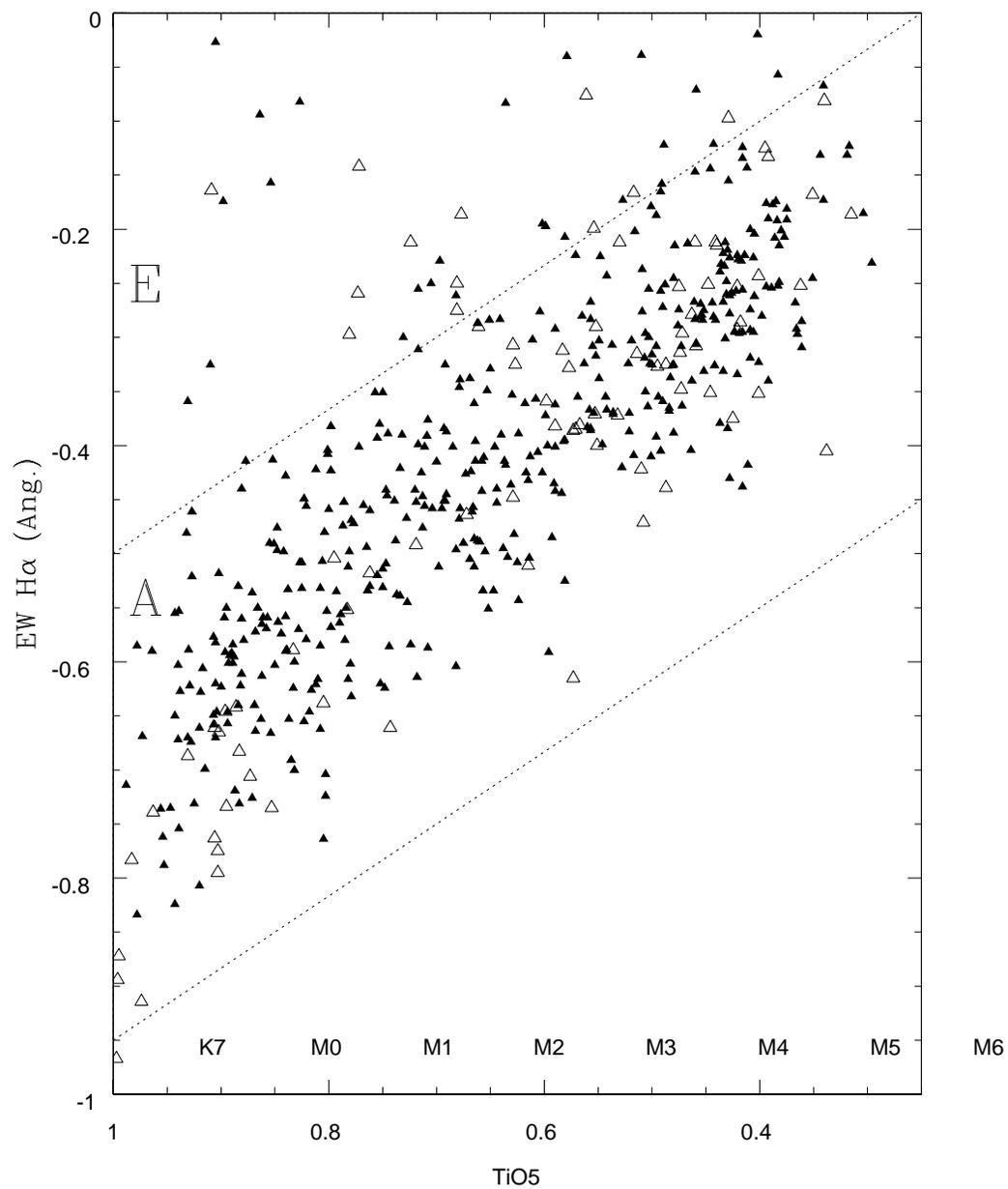}
\caption{H$\alpha$ absorption as a function of TiO5 (spectral type).
Symbols are as in Figure~\ref{fig-tio5ha1}.  \label{fig-tio5ha2}}
\end{figure}

The primary indicator of chromospheric activity in M dwarfs 
is H$\alpha$ emission.
In Figure~\ref{fig-tio5ha1} we plot the 
equivalent width of H$\alpha$
as a function of TiO5 (spectral type/effective temperature).  
Figure~\ref{fig-tio5ha2} is an expanded view of the absorption
(negative) equivalent width portion of Figure~\ref{fig-tio5ha1}.
Active stars known to be short-period binaries
(Tables~\ref{table-sb2} and~\ref{table-sb1}) are shown as
open circles while members of the VC sample are shown 
as solid triangles.  Additional stars from this paper which
are not in the VC sample are shown as open triangles.  
To aid discussion, we have labelled five groups (A-E).

{\it Group A:}  The majority of M dwarfs show H$\alpha$ absorption,
with larger average absorption equivalent widths at early spectral types.
H$\alpha$ absorption does not necessarily indicate an absence of 
chromospheric activity, as discussed by
\citet{cm79} and \citet{cg87}. These authors showed that the
presence of a weak to moderate chromosphere induces H$\alpha$ absorption in 
M dwarfs, which enhances the photospheric absorption.  Only
a rather strong chromosphere will produce H$\alpha$ emission.
Thus, M dwarfs with the strongest H$\alpha$ absorption probably
have moderate chromospheres, while those with no chromosphere
will exhibit only weak (or perhaps no) absorption -- see
discussion of Group E. 
The observed width of the absorption sequence in Figure~\ref{fig-tio5ha2} is
larger than our measuring uncertainty, and therefore probably reflects real 
star-to-star variations at a given effective temperature.  These variations
are presumably correlated with the level of chromospheric activity. 
Also, the fraction of stars per spectral type bin 
which are in Group A decreases towards later 
spectral types, which is probably related to the increase in the lifetime 
of strong chromospheric activity for lower mass stars (see discussion in
Section~\ref{age}).

{\it Groups B and C:}  Most of the early (M0-M2.5) dwarfs  
with emission, including 
all the early-M active binaries (BY Dra class) in the VC sample,
cluster near the upper envelope of observed activity.  We have marked
these dwarfs as Group B.  There is a striking lack of early-M dwarfs
with weaker emission, indicated as Group C.  
This behavior was noted by \citet{hm89}, who point out a similarity with the
Vaughn-Preston gap in the CaII emission of G dwarfs \citep{vp80,hsdb96}.  
Such a gap can be interpreted either as a stellar population (age) 
or a stellar chromosphere effect.  In the former case, one assumes that
age and H$\alpha$ emission have a simple relationship, and
the distribution reflects a lack of intermediate age
(Group C) stars compared to younger (Group B) and older (Group A)
stars.  In the latter interpretation, 
some aspect of chromospheric physics makes it unlikely that
a star is observed with intermediate-strength emission lines.  
For example, stars may remain in the Group B state for 
a long time, then rapidly evolve through the Group C region.

{\it Group D:} The observed distribution  of H$\alpha$ equivalent
width changes at TiO5 $\sim 0.5$ (spectral type M2.5/3).  
Amongst earlier spectral types, emission is limited to 
$\sim 2.5$\AA\ above the mean level of absorption; for the 
mid-M (M3-M5.5) dwarfs, the emission-line equivalent
widths are distributed roughly uniformly between 0\AA\ and a
maximum value of 6-10\AA. This change in properties was 
also noted by \citet{hm89} on the basis of data for fewer stars. 
Little structure is evident, although there is a suggestion of
a clump at H$\alpha \sim 4.5$\AA, TiO5$ \sim 0.4$.
According to the empirical mass-spectral type relations
derived by \citet{km94}, spectral type M3 corresponds to $0.26 M_\odot$,
while M2 corresponds to $0.37 M_\odot$.  This range
encompasses the mass where M dwarfs are expected to 
become fully convective ($\sim0.3 M_\odot$, \cite{bur97}). 
The change in the distribution 
of activity strengths may be related to changes
in the magnetic dynamo generation mechanism and/or the
atmospheric structure amongst fully-convective stars \citep{flagmtg}.

{\it Group E:}  A few dM stars have unusually weak H$\alpha$
absorption.  They are sometimes called dM(e) stars, since
they appear to be intermediate between the dMe and dM stars.  
Two different states contribute to this group.
Some stars may have extremely weak, or possibly no, chromospheres.
The absorption line, or lack thereof, then simply reflects the
weak photospheric absorption which disappears at later types.
Others have moderately strong chromospheres that have begun to
produce enough emission to fill in the absorption line, but
not enough to produce a full-fledged emission line.
Examples of both types are known \citep{byrne93,dhmp94}.
Two of the stars in Group E, GJ 1062 (NN 647) and LHS 64 (NN 3308) are old, metal-poor
M subdwarfs \citep{g97}, and as such are likely to have weak,
if any, chromospheres.  Further study of Group E stars may
be useful to characterize this rare, inactive state amongst
M dwarfs.  

An interesting question raised by our analysis is whether
there is an evolutionary pattern through the groups we
have identified.
Thus, for example, an early type M dwarf might start out
quite active (group B).  As the chromosphere begins
to weaken with age, the H$\alpha$ emission may fade rapidly
according to the particulars of the line radiative transfer
leading to a quick passage through Groups C and E.  Further
chromospheric weakening may lead to a slower evolution
from the top (less absorption) of Group A to the bottom
of Group A, and finally, as the stars age even further
(approaching the age of the Galactic disk?) the chromosphere
begins to disappear altogether, with the evolutionary
path progressing back through Group A to an endpoint in Group E.
For later type stars, it may be that the radiative transfer
in the H$\alpha$ line is such that it is easier to drive the
line into emission.  Thus those stars start, and remain, in
Group D throughout most of their evolution, at least to
their current ages.  We are pursuing these ideas by searching
for other chromospheric indicators that would allow us to distinguish
between moderate activity (filled-in H$\alpha$) and no
activity (weak or no H$\alpha$ absorption), and by developing
chromospheric models that predict the behavior of
the H$\alpha$ (and other) emission lines depending on
parameters that characterize the strength of the chromosphere
(Hawley et al., in preparation).

\begin{figure}
\epsscale{0.8}
\plotone{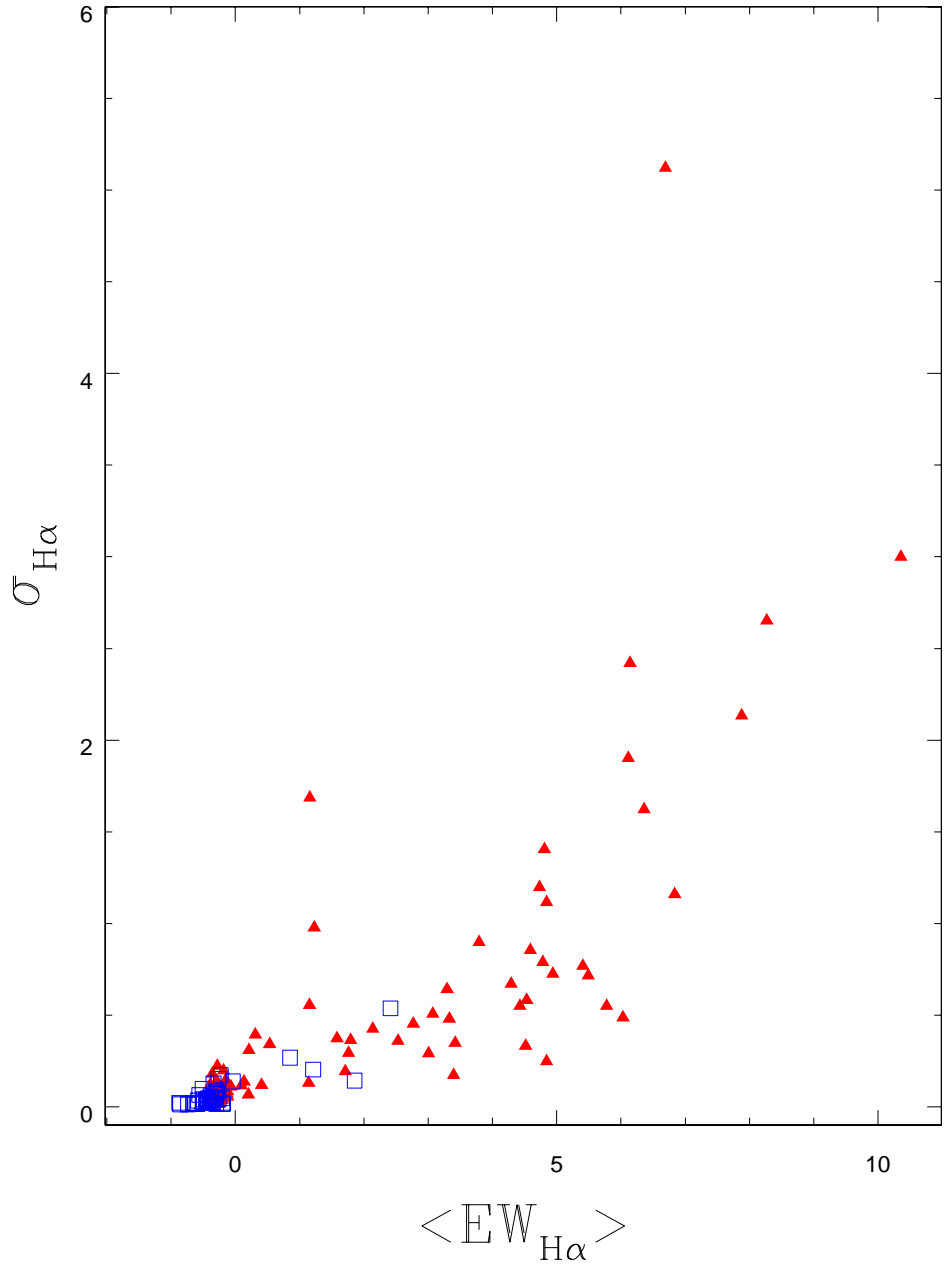}
\caption{The observed standard deviation ($\sigma_{H\alpha}$) in
the H$\alpha$ equivalent width as a function of EW$_{H\alpha}$ for
dMe stars with at least 4 measurements.  Stars
with EW$_{H\alpha} > 5$\AA exhibit significantly larger scatter.
\label{fig-hasig}}
\end{figure}

We obtained monitoring observations of the nearest M dwarfs 
as part of this program.  These allow us 
to examine variability in the emission line strength. 
Figure~\ref{fig-hasig} 
plots the observed standard
deviation in equivalent width as a function of emission line 
strength for stars with at least 4 observations.  For 
dMe dwarfs with H$\alpha \lesssim 5$\AA\ (including
all early M dwarfs and the weaker mid-M dwarfs), variations
are $\lesssim 0.5$\AA, with typical variations at the
15\% level.  In contrast, stars with stronger emission
are significantly more variable, with variability exceeding 30\%, 
corresponding to changes of several Angstroms in equivalent width.

\begin{figure}
\epsscale{0.8}
\plotone{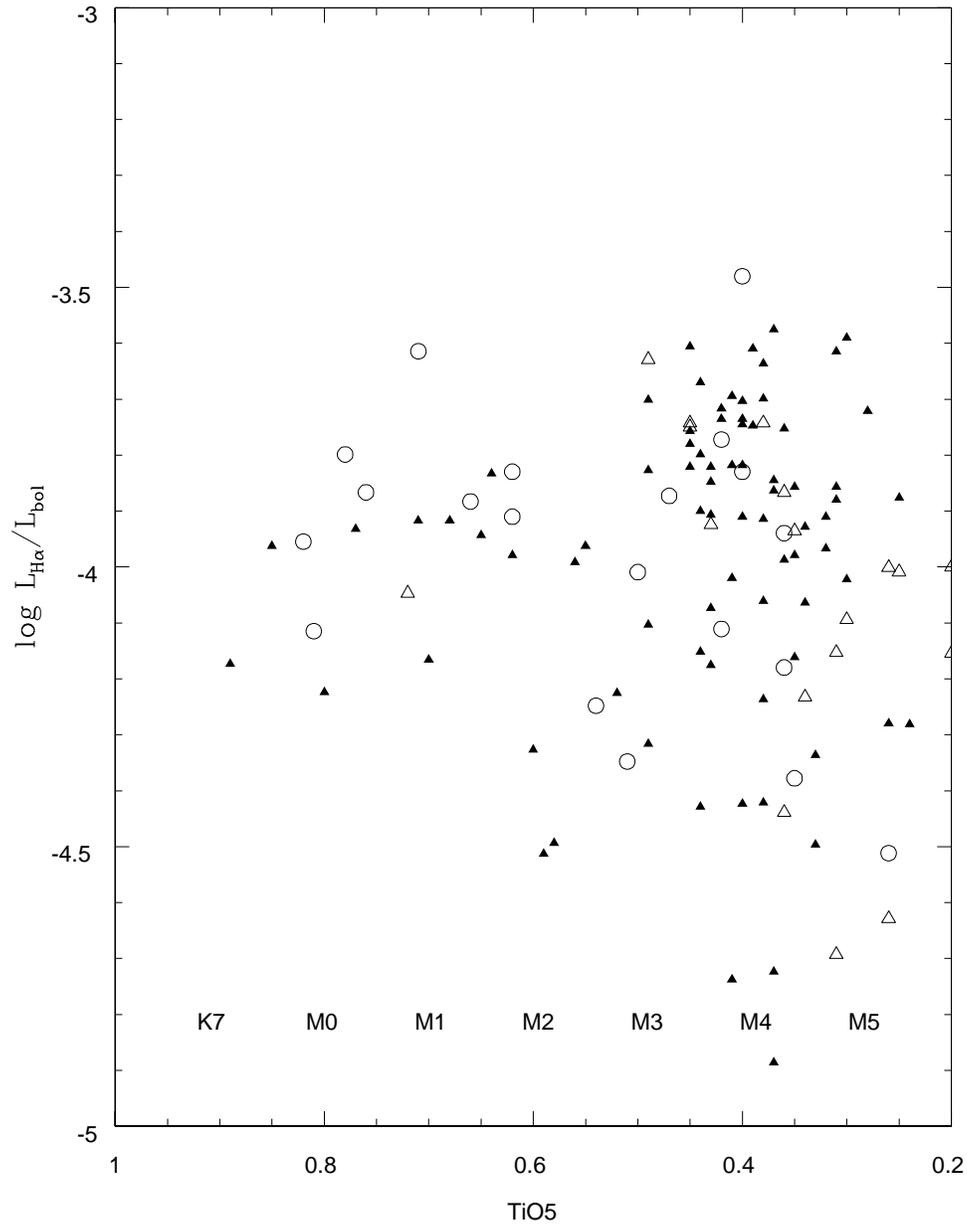}
\caption{The chromospheric activity level expressed as the ratio of
the H$\alpha$ luminosity to the bolometric luminosity. 
Symbols are as in Figure~\ref{fig-tio5ha1}.
\label{fig-tio5lrha}}
\end{figure}

As discussed extensively elsewhere (e.g. \cite{rhm95}), 
equivalent width measurements are
useful in segregating stars within a relatively limited range of effective
temperature, as in defining the groups in Figure~\ref{fig-tio5ha1}, but
do not provide a good measure of the absolute level of activity,
due to variations in the continuum flux as the effective temperature
changes. We have therefore used the 
relations defined in Section~\ref{lines}
to transform our equivalent width
measurements to line fluxes, and formed the ratio of the
line luminosity to the bolometric luminosity -- an absolute 
measure of the activity strength for each star.  Figure~\ref{fig-tio5lrha} 
plots the result for H$\alpha$ as a function of TiO5 (equivalently colour, 
spectral type, effective temperature, mass).  The Group B stars
are clearly evident at spectral types earlier than M2, while the
Group D stars comprise the cooler half (smaller TiO5) of the diagram.  It is
interesting that binaries (open circles) are found scattered throughout 
the latter group, with no clear clustering at the most active levels.  
The discussion of the groups given above is applicable also to this figure, 
with the Group B stars showing high and rather uniform activity and
some evidence for binaries having stronger activity, while the Group
D stars have a much larger scatter.  The mean level of 
log (L$_{H\alpha})$/L$_{bol}$) 
is $\sim$ -4.0, in agreement with the
value we found for the Hyades \citep{rhm95} and in our lower resolution
survey of these same field stars (Paper II).  Figure~\ref{fig-tio5lrha}
makes it clear however, that the scatter among the Group D stars (later types)
is not uniform.  There are more stars that show activity above the mean
level (between -3.5 and -4), but the stars below the mean level have a
larger range (between -4 and -5).  With reference to our previous speculative
discussion about chromospheric evolution, this could be evidence that 
the younger, most active stars cluster around 
log (L$_{H\alpha}$/L$_{bol}$) $\sim -3.8 \pm 0.2$, while the older,
less active stars spread out according to their age and initial
activity level.  It is notable that the difference in scatter
between Group B and Group D is still very  striking, which may mean that some
change in the dynamo generation at the fully convective boundary is
also required (along with evolution and the particulars of H$\alpha$ line
transfer) for a full physical explanation of these phenomena.

\begin{figure}
\epsscale{0.8}
\plotone{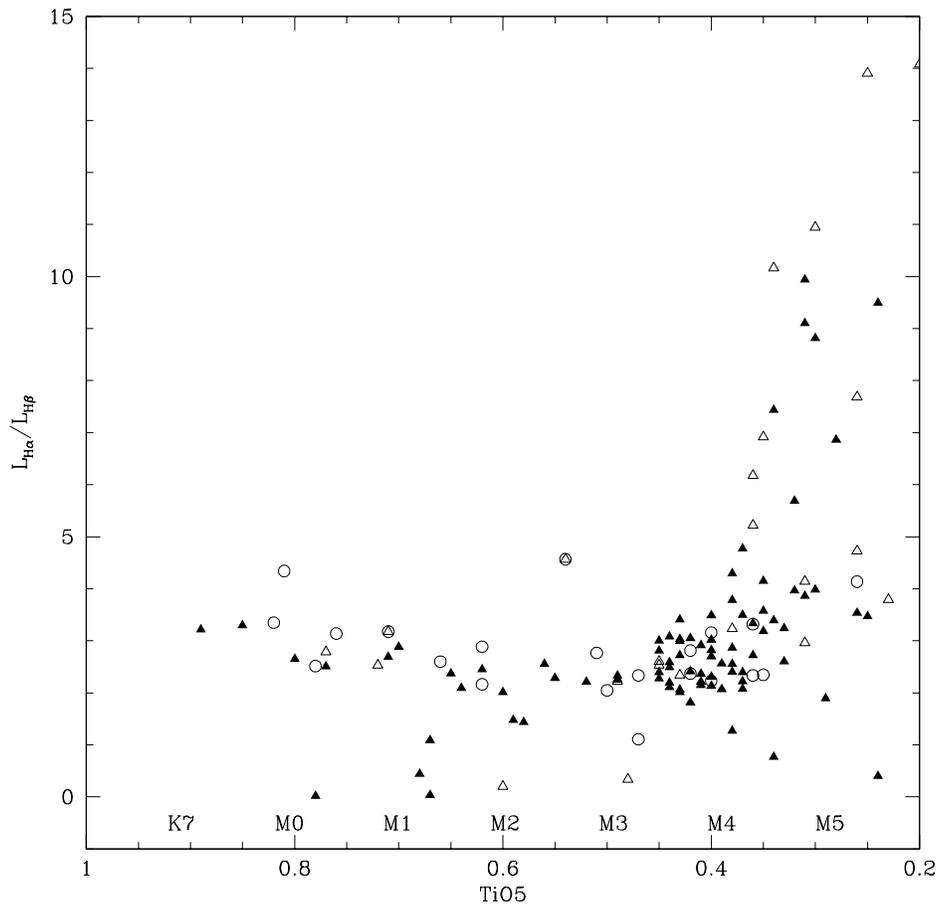}
\caption{The ratio of the H$\alpha$ to H$\beta$ emission. 
Symbols are as in Figure~\ref{fig-tio5ha1}.
\label{fig-tio5rbalmer}}
\end{figure}

Figure~\ref{fig-tio5rbalmer} shows the Balmer decrement, the
ratio of H$\alpha$ to H$\beta$ flux.  A typical value in the Hyades
is $\sim$ 4, somewhat larger than the mean found here of $\sim$ 2.5 (which
does agree with our result in Paper II).  There is a slight 
downward trend toward later types among the Group B stars and again
a notable increase in the scatter, particularly toward larger values
of the decrement, among the Group D stars.  It may be significant
that this scatter is apparently restricted to the later spectral
types among Group D.  The scatter is reminiscent of that seen in
Figure~\ref{fig-hasig} in $\sigma_{H\alpha}$.  There are 8 stars shown
with $\sigma_{H\alpha} > 1.5$ kms$^{-1}$,  and three of these (NN 1702,
1724, 1326) have the three highest values of the Balmer decrement, while
four of the remaining five (NN 204, 230, 1398, 1934) have decrements 
well above the average, between 4 and 10.  A very strong Balmer decrement
and significant variability probably indicate that we have observed
flaring activity
in these stars; the decrement actually $decreases$ in flares on
earlier M dwarfs (e.g. AD Leo, \citet{hp91}), but is known
to increase strongly during flares on later M dwarfs (e.g. VB 10,
\cite{herbig}; 2MASSJ0149, \cite{l99}).  The final star, LHS 1723 (NN 855)
is anomalous in both figures, lying at $\sim$ (1,1.7) in Figure~\ref{fig-hasig}
and at $\sim$ (0.4,1.2) in Figure~\ref{fig-tio5rbalmer}.  This star
has very weak H$\alpha$ emission compared to H$\beta$ and also shows
far more variability than other weak H$\alpha$ emitters.
It will pose an interesting challenge for chromospheric models.

\begin{figure}
\epsscale{0.8}
\plotone{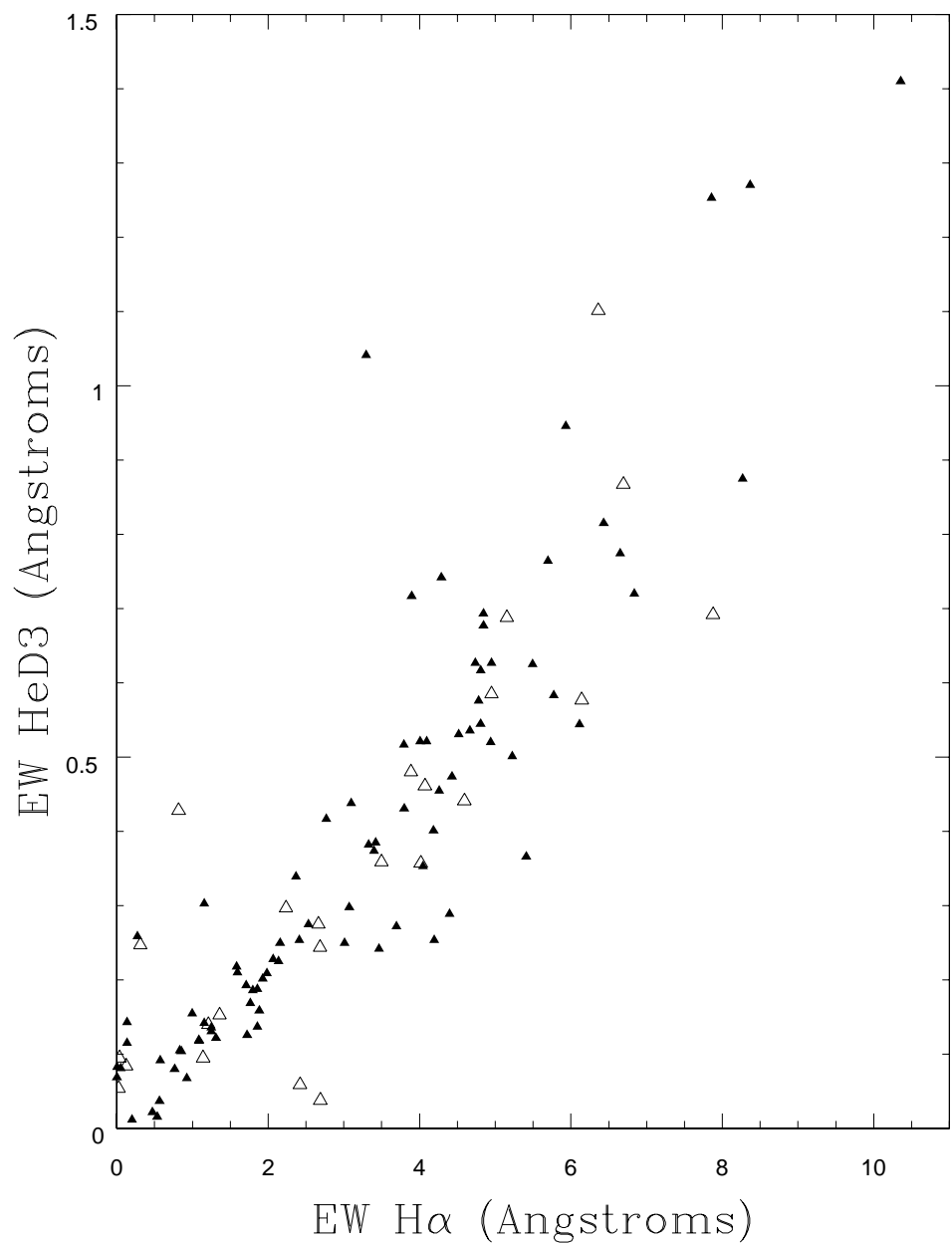}
\caption{He D3 emission as a function of H$\alpha$ emission.  
Symbols are as in Figure~\ref{fig-tio5ha1}.  
\label{fig-hahed3}}
\end{figure}

Finally, our observations comprise the most extensive available set
of Helium D3 ($\lambda 5876$\AA) emission line data for active M dwarfs. 
Figure ~\ref{fig-hahed3} 
shows that there is a strong correlation between the He D3 and
H$\alpha$ emission equivalent widths, with the slope of a simple linear 
fit $\sim$ 0.12.  Alternatively, the flux in the He D3 line may
be predicted from the H$\alpha$ flux by $F_{HeD3} = 0.048 \times F_{H\alpha}$.
The good correlation suggests that both features probably 
originate from the same region within the chromosphere.  This is in contrast
with previous observations and modelling of Ca II, Mg II and H$\alpha$ 
emission in M dwarfs which implied that the emission in those lines was coming
from different chromospheric regions \citep{mf94,gwl82}.
An extensive study of He D3 in G and K stars by \citet{saar}
suggested formation in the upper chromospheres of those stars,
and good correlation of the He D3 flux with those of Ca II and C IV
(similar to our result here for H$\alpha$).  \citet{andretta} have
outlined a method using He D3 observations to infer the filling factor 
of active regions in F and G stars.  It would be of great interest 
if such a technique was applicable also for M dwarfs.

\subsection {An age-activity calibration for M dwarfs\label{age}}

Calibrating age-dependent relations obviously demands 
stars with well determined ages.  In the Galactic disk, open clusters
give an independent age determination, and observations
of chromospheric activity levels of cluster members
provide the calibration.  The traditional approach, as
in \citet{barry}, is to determine a relationship
between the activity level (e.g., equivalent width or other
measure) and the age.  By applying this law to each field star,
unique ages can be derived.  While we have proposed above a
possible qualitative evolutionary sequence that might 
eventually allow
a unique age determination for each M dwarf in our sample,
we are far from certain that our proposal is correct, and
have not even begun to determine a calibration.  Also,
observations of clusters show that there is some spread
in activity at a given age.  Moreover, as demonstrated in 
Figure~\ref{fig-hasig}, variability is clearly present in 
individual M dwarfs, complicating any effort to determine
an exact activity level.  Finally, the appearance of
Figure~\ref{fig-tio5ha1} suggests potential inconsistencies in
applying this method: If the lack of Group C stars is interpreted 
as a lull in star formation, where is the corresponding
gap in Group D?  If the concentration of stars in Group B
is a burst of star formation, why is there no corresponding
concentration at the top of Group D?  

\begin{figure}
\epsscale{0.8}
\plotone{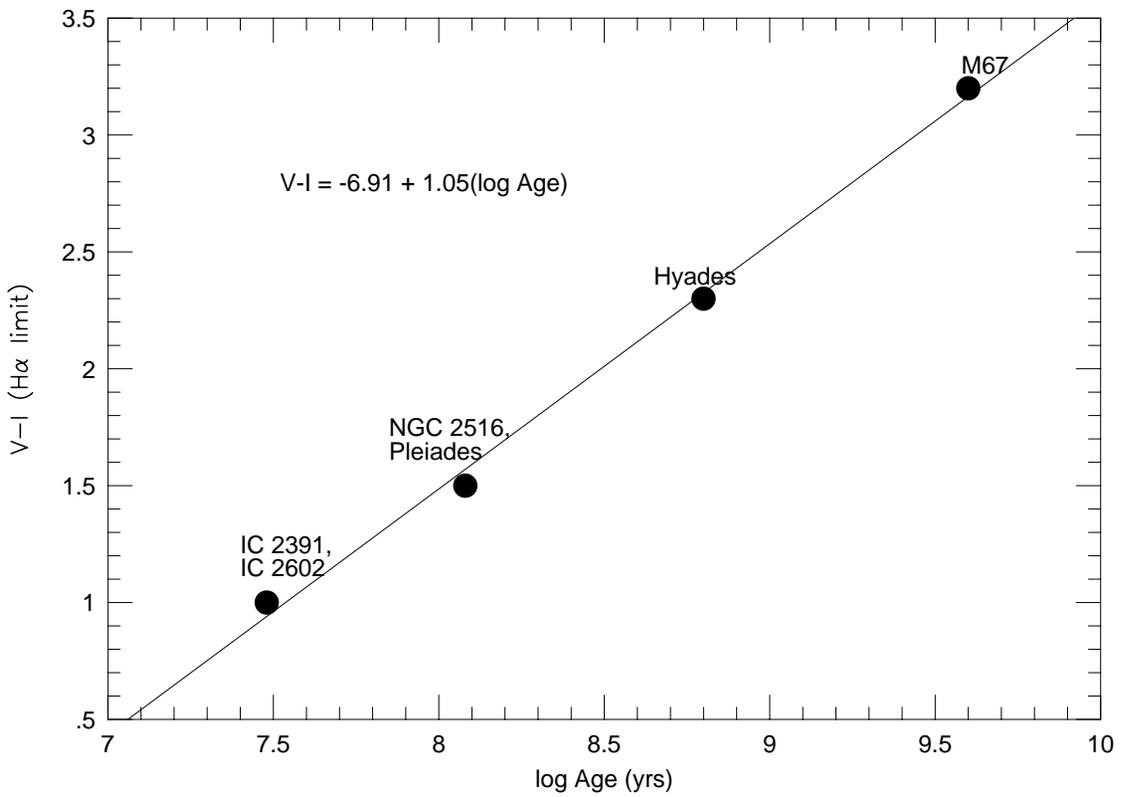}
\caption{Calibration of $V-I_C$ at the H$\alpha$ limit
as a function of cluster age, from Hawley et al. (1999).  See
text for discussion.
\label{fig-vi-age}}
\end{figure}

Hawley et al.'s (1999) analysis of M dwarfs in open clusters
suggests a different approach.  While the activity levels of stars 
in a given cluster exhibit considerable scatter, 
there is a well-defined $V-I_C$ colour at which activity becomes
ubiquitous. All stars redder than this colour are
dMe (defined as $EW_{H\alpha} \ge 1.0$\AA), while the bluer stars 
are dM without emission.  This effect was commented on originally by
\cite{s94} for the Pleiades.  Observations currently
exist for M dwarfs in six clusters:
IC2602 \& IC2391 (30 Myrs, Barnes et al. 1999),
NGC 2516 \& the Pleiades (125 Myrs, Stauffer, Schultz \& Kirkpatrick 1998), 
the Hyades (625 Myrs, Perryman et al. 1998),
and M67 (4.0 Gyrs, Dinescu et al. 1995).  Hawley et al. (1999) have used those
observations to determine the relationship between the `H$\alpha$ limit' 
colour and the age in years, 
\begin{equation}
\label{equation-age-vi}
V-I_C = -6.91 + 1.05(\log Age )
\end{equation}
as shown in Figure~\ref{fig-vi-age}. 

We can use this correlation to transform an
observed distribution of chromospheric activity as a function of $V-I_C$ colour to 
an estimate of the cumulative age distribution.  We define $f_{dMe}$ as the
fraction of dMe dwarfs at a particular colour, $(V-I_C)_i$, 
\begin{displaymath}
f_{dMe} \ = \ N ({\rm dMe} ) / N ({\rm dM + dMe}), \ (V-I_C) = (V-I_C)_i \pm \delta
\end{displaymath}
This ratio provides an estimate of the fraction of stars younger than age
$\tau_i$, where, from Equation~\ref{equation-age-vi}, 
\begin{displaymath}
\log \tau_i \ = \ ( (V-I_C)_i + 6.91)  / 1.05
\end{displaymath}
Thus, for example, the fraction of dMe dwarfs at $V-I_C$ = 1.0 is directly proportional to the
relative number of stars which have formed in the last $\sim 3 \times 10^7$ years, while
$f_{dMe}$ at $V-I_C$ = 2.6 corresponds to the relative number of stars with
ages of less than 1 Gyr.  In this manner, we can determine $f_{dMe}$ for a range of 
$V-I_C$ colour, and map the cumulative star formation history of the Galactic disk. 

\subsection{Star Formation History of the Galactic Disk}

\begin{figure}
\epsscale{0.8}
\plotone{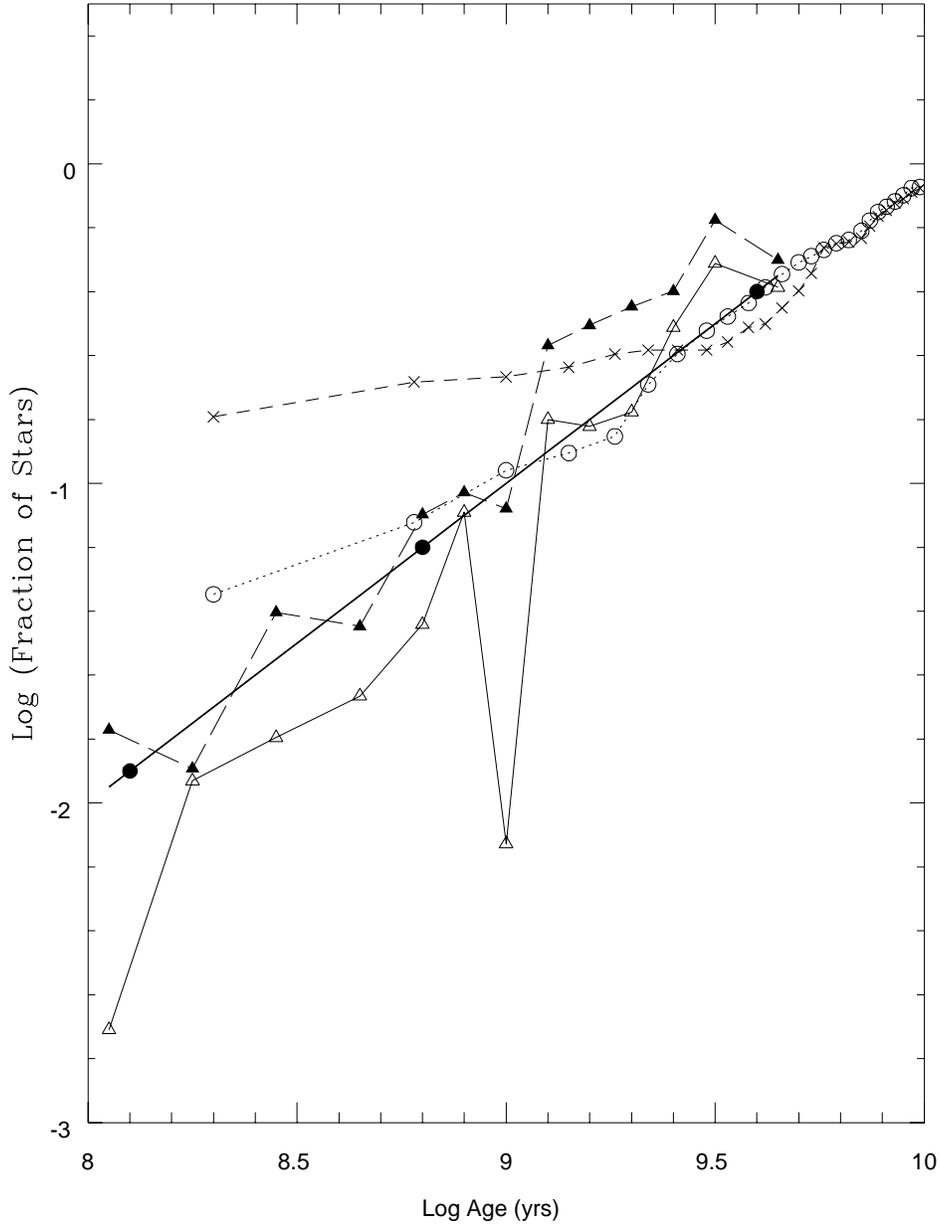}
\caption{The star formation history of the Solar Neighbourhood
(assuming a Galactic disk age of 10 Gyr).
Our M dwarf results are shown for the W-weighted (solid line, open triangles) 
and unweighted (long-dashed line, filled triangles) analyses.  The thick 
solid line, slope unity, illustrates the expected distribution for a constant 
star formation rate. The closed circles on this line
mark the ages of the calibrating open clusters.
Barry's (1988) G dwarf results (short-dashed line, crosses) 
are also shown together with the results from the \cite{rp00} G dwarf 
analysis (dotted line, open circles).
The M dwarf results do not show the excess of young stars suggested by 
the former
analysis and also fail to match the details of the latter.
\label{fig-sfhist}}
\end{figure}

We have applied the analysis technique outlined in the previous section to
data for the M dwarfs in the VC sample.  For stars lacking $V-I_C$ measurements, 
we use the TiO5 index to estimate the colour.
The results are plotted in Figure~\ref{fig-sfhist}, giving
$f_{dMe}$ as a function of the inferred age. 
A constant star formation rate (i.e. $f_{dMe}$ directly proportional to age)
is shown for reference as the thick solid line with slope unity in this figure. 
Two different ways of computing the fraction of dMe stars are illustrated.
The long-dashed line (connecting solid triangles) gives the fraction found 
by weighting all stars equally.  The thin solid line (connecting open triangles)
shows the fraction found by weighting each star by the inverse of its 
W velocity, 
as proposed by \citet{w74,w77}.  The latter approach allows for the
increased scale height of higher velocity (older) stars, and consequent
shorter residence time in the Solar Neighbourhood. However, this method 
adds statistical noise by placing significant (undue?) weight on a small
number of high velocity stars. In both cases we exclude the SB1 and SB2 
(short period binary) systems, since the activity in those stars may be
influenced by other effects than age.

Figure~\ref{fig-sfhist} indicates that the overall star formation
history is broadly consistent with a constant star formation
rate.  The major feature notable in the distribution is 
a step at $\sim$ 1 Gyr which, if taken at face value,
would indicate that $\sim10$\% of the 
local disk stars formed in a burst at that time.
However, this feature is unlikely to be real, since it corresponds
with the Group B/D transition in Figure~\ref{fig-tio5ha1} which 
we discussed at length in Section~\ref{activity}.
Such a distinct change in activity properties is unlikely to be well described 
by a simple linear relation.  Data for clusters with ages between the 
Hyades (0.6 Gyr) and M67 (4 Gyr) would be useful in confirming whether 
this feature has an astrophysical, rather than evolutionary, origin.

The other characteristic of the star formation history 
illustrated in Figure~\ref{fig-sfhist} is a slight deficiency
in the number of young ($<1$ Gyr) stars relative to
the number of older ($>1$ Gyr) stars.  At ages
of less than 1 Gyr, both the weighted and unweighted solutions
match the expectation of constant star formation (slope unity), 
with slopes of  $0.92 \pm 0.16$ and $0.95 \pm 0.50$ respectively.  
For the full age distribution, however, we derive best-fit slopes of
 $1.14 \pm 0.09$ for the
unweighted solution and $1.38 \pm 0.19$ for the weighted
solution. This result implies 
a slight decrease in the star formation rate 
in recent times. However, this could reflect incompleteness in
the VC sample which, since it is derived primarily from  proper 
motion surveys, may be deficient in young, low space-motion dwarfs.

Figure~\ref{fig-sfhist} also compares our analyses with the star formation
histories proposed from G dwarf studies by \citet{barry} and \citet{rp00}. The 
poor time resolution at large ages, coupled with the lack 
of calibrating clusters and the relatively small size of the VC sample, 
limits the utility of comparisons at older ages ( $>3$ Gyrs). Nonetheless,
it is encouraging that all of the models predict similar fractions
of stars younger than 4 Gyrs. 
At younger ages, the M dwarf data are better calibrated. There is no
evidence from our analysis for the substantial numbers of young G 
dwarfs ascribed by \citet{barry} to a recent burst of star formation.
Similarly, our data fail to match the details of the \citet{rp00} analysis.
Indeed, \citet{rp00}'s ``Burst A'' of young ($\lesssim 0.5$ Gyr) G dwarfs is 
accompanied by a deficiency of M dwarfs, while their ``AB Gap'' at 1-2 Gyrs
corresponds to an apparent excess of M dwarfs (although, as we noted above, we
believe this feature reflects a deficiency in our analysis method rather
than a burst of star formation). 

The ``AB'' gap found by \citet{rp00} corresponds to the Vaughan-Preston gap
\citet{vp80}. As noted above (Section~\ref{activity}), \citet{hm89} have 
suggested that the sparse number of early type M dwarfs with weak emission 
(our Group C) might be an analogous
feature. However, observations of Hyades M dwarfs show that the
H$\alpha$ limit corresponds to TiO5$\sim$0.55 in that cluster. 
Thus, the weak emission 
M dwarfs in Group C must have ages of less than the age of
the Hyades (0.6 Gyrs) -- i.e. ages which match
\citet{rp00}'s ``Burst A''. As \citet{sdj91} and 
\citet{rp00} have discussed, one needs 
either a complicated G-dwarf age-activity relation
to match a constant star formation history, or a 
complicated star formation history to save the
simple G-dwarf age-activity relation.  Given the lack of 
agreement between our M dwarf analyses and the G dwarf analyses, we believe
that the complicated G-dwarf age-activity relation explanation is
favored.  The existence of a large spread in activity in the
coeval G dwarfs of M67 \citep{giampapa_m67} 
and the large spread in rotation rates in stars
of young clusters \citep{barnes} suggests that 
any age-activity relation is complex, with stochastic star to star variations.
Indeed, M dwarfs show similar behaviour, with a smattering of dMe dwarfs
bluer than the H$\alpha$ limit in some clusters. Binarity may well be
a contributing factor at both spectral types. 

Other measures of the recent star formation history have
been made without reference to chromospheric activity. 
\citet{x00} have used Hipparcos color-magnitude diagrams 
to derive the star formation history of the Solar Neighbourhood
within the last 3 Gyrs.  They find an oscillatory component
of star formation with a period of 0.5 Gyr superposed on an
underlying constant star formation rate.  Their Figure 4 indicates that
they see roughly 2.5 times as much star formation from 
$\sim 1.6-2.6$ Gyr as at $\sim 0.3 - 1.0$ Gyr.  It is suggestive
that this is similar to the deviations from constant
star formation seen in our Figure~\ref{fig-sfhist} --
that is, the possible deficiency of young stars.  Neither
our data nor \citet{x00} show the lull in star formation between
1-2 Gyrs seen by \citet{rp00}.  

Our present analysis is only a first step towards using 
M dwarfs as probes of Galactic star formation.  
Additional observations of M dwarfs in
clusters, particularly clusters older than the Hyades, are
required to improve the age-activity calibration during the
important 1-4 Gyr time period.
Further observations in clusters and the field are needed to quantify
the significance and range of star-to-star variations, and
(with sufficient resolution) to provide additional information
on the evolution of activity at a given spectral type, both
among the Group B and Group D stars.
Finally, a nearby star sample which is both unbiased kinematically
and complete over a larger volume will provide improved statistics for
the local field stars.  We are currently developing such a
sample in a project undertaken under the auspices of the NASA/NSF NStars program
(Reid \& Cruz, 2001). 

\subsection{Brown Dwarfs}

The ability to constrain the ages of individual M dwarfs can be 
used to set limits upon the ages of brown dwarf candidates which
happen to be in binary systems with M dwarf primaries.  This allows
an independent test of the predictions of theoretical model isochrones.
At the present time, two L dwarfs and two T dwarfs are known to be companions
of nearby M dwarfs.  In addition, \citet{gl569} have shown that Gl 569 A 
is the companion to a binary system, Gl 569Ba and Bb. 
In Table~\ref{table-bd}, we use the
presence or absence of H$\alpha$ emission in the primary, together
with Equation~\ref{equation-age-vi}, to constrain the age of each system.  
The discovery groups have also discussed the limits that chromospheric activity
places on the age.  In particular, it should be noted that G 196-3
is a very active M2.5 dwarf, so \citet{g1963b} argue that
it is likely to be younger than our limit.  Based on the lack of emission in 
Gl 229A, as well as the fact that no rotational broadening is detectable, a
the young age of 30 Myr advocated by \citet{badgl229b} is ruled out --- 
indicating that the synthetic spectra from model atmospheres 
are not yet adequate for age determinations.  
Continued discoveries of L and T dwarf companions to nearby M dwarfs
may be expected from both dedicated searches and the 2MASS, DENIS,
and SDSS surveys, and improved age determinations will aid in the 
usefulness of those data.

Our star formation results also apply to the modelling of
the substellar mass function.  Because age and mass are
not directly observable for isolated field brown dwarfs, 
a mass function can only be derived through simulations.
Reid et al. (1999) have modelled L dwarf detections from DENIS
and 2MASS, assuming a
constant star formation rate for the Solar Neighbourhood.
Our results show that this assumption is largely consistent with
our analysis of 
the local low-mass stars, and therefore is likely to 
be appropriate for the local brown dwarfs as well.  However, if there is
a slight deficiency in the local density of very young stars (and
brown dwarfs), then the numbers of L- and T-type brown dwarfs that we 
observe is a smaller proportion of the number that actually exist (because
they are easier to observe when they are younger/brighter).
In this case, the Reid et al. analysis will underestimate both
the slope of the mass function and
the total number of brown dwarfs in the Galactic disk.  A reliable
estimate of the local star formation history is critical to determining
the contribution made by brown dwarfs to the mass of the Galactic disk.

\section{Summary}

We obtained high resolution spectra of a large
sample of nearby M dwarfs to supplement our earlier 
moderate resolution survey.  The new measurements include
accurate radial velocities, H$\alpha$ emission and absorption
equivalent widths, and simultaneous observations of several 
other chromospheric lines, including Helium D3.  We discussed
several groupings of stars based on their H$\alpha$ activity and 
speculated on the nature of the chromospheric evolution, including 
differences between early (M0-M2.5) and mid (M3-M6) dwarfs.

We also argued that the relative numbers of dM and dMe
stars as a function of spectral type, together with an
age-activity relation calibrated for M dwarfs, may be used to 
probe the star formation history of the Galactic
disk.  We found no evidence for the various recent 
bursts and lulls in star formation history inferred from 
G-dwarf activity results.  This suggests that the G-dwarf
age-activity relationship is complex, and must be better
understood before attempting to use it to derive the Galactic
star formation history.  Finally, we illustrated the use of M dwarf 
activity levels to constrain the ages of binary systems
with brown dwarfs.  

Additional study of active M dwarfs in 
open clusters and the field will allow a
better calibration of the M dwarf age-activity relationship,
and more accurate determination of the local star formation
history, which is crucial for understanding the substellar
mass function.  When coupled with chromospheric models,
we can also hope to attain a better understanding of the
underlying physical processes (evolution, atmospheric structure, 
dynamo generation) that ultimately control the empirical
behavior we observe.

\acknowledgments

We gratefully acknowledge the assistance of Skip Staples and the other 
telescope operators at the Palomar 60-in. telescope.
Jim McCarthy gave helpful assistance with the echelle hardware and software.
We would like to thank the staff of Palomar Observatory and Jim McCarthy
for their work on making the new cross-dispersing prisms a reality.
Tokovinin provided an electronic version of his radial velocity data.  
JEG was partially supported by Greenstein and Kingsley Fellowships.
SLH was partially supported by the NSF through Young Investigator
award AST 94-57455.  
This research has made use of the Simbad database, operated at
CDS, Strasbourg, France.

\clearpage

\pagestyle{empty}
\begin{deluxetable}{lrcrr}
\tablecaption{Radial Velocity Comparison}
\tablewidth{0pt}
\tablenum{1}
\tablehead{
\colhead{Source} & 
\colhead{N} &
\colhead{$\sigma_{ref}$} &
\colhead{$<v_{rad}>$ } & \colhead{$\sigma$}  \\
\colhead{} &\colhead{} & \colhead{kms$^{-1}$} &
\colhead{kms$^{-1}$} & \colhead{kms$^{-1}$}
}
\startdata
 Bopp and Meredith (1986) & 17 &1-2 & -1.07 & 3.58 \\
Delfosse {\sl et al.} (1998) & 93 &0.5 & 0.04 & 2.38 \\
Delfosse {\sl et al.} (1998) & 90$^*$ &0.5 & 0.12 & 1.25 \\
 Stauffer and Hartmann (1986) & 115 &0.9 &-1.27 & 1.99 \\
 Tokovinin (1988,1992) & 51 & 0.5 &-0.75 & 1.38 \\
 Upgren and Caruso (1988)& 64 & 1 &-0.11 & 1.58 \\
 Upgren and Harlow (1996) & 25 & 1.2& -0.97 & 1.83 \\ 
\enddata
\tablecomments{
$^*$ excluding the three stars discussed in the text}
\end{deluxetable}

\begin{table}
\dummytable\label{table-vcomp}
\end{table}

\pagestyle{empty}
\begin{deluxetable}{rrrrrrrrrr}
\tablewidth{0pt}
\tabletypesize{\scriptsize}
\tablecaption{Radial Velocities}
\tablenum{2}
\tablehead{
\colhead{NN} & \colhead{Name} & \colhead{Julian Date} & \colhead{$v_{rad}$} &
\colhead{Julian Date} & \colhead{$v_{rad}$} &\colhead{Julian Date} & \colhead{$v_{rad}$} 
& \colhead{ $\langle V_{rad} \rangle$} & \colhead {$\sigma$}
}
\startdata
    7& Gl   2      &  2450008.6917 &     0.4& & & & &    0.4&   \nodata \\
   11& Gl   4  A   &  2449561.8691 &     3.3& & & & &    3.3&   \nodata \\
   12& Gl   4  B   &  2449561.8780 &    -1.6& & & & &   -1.6&   \nodata \\
   19& GJ 1002     &  2450085.6899 &   -39.5 &  2450285.9580 &   -40.7 &  2450473.6302 &   -39.5&  -40.1&    0.6 \\
     &             &  2450668.9841 &   -40.8 \\
   38& G242-048A   &  2449559.9545 &   -13.2& & & & &  -13.2&   \nodata \\
   41& GJ 1005 B   &  2450009.7126 &   -23.3 &  2450086.6392 &   -26.8 &  2450304.9421 &   -27.3&  -25.5&    1.7 \\
     &             &  2450473.6410 &   -23.0 &  2450668.9975 &   -26.7 &  2450670.8852 &   -26.1 \\
\enddata
\end{deluxetable}

\begin{table}
\dummytable\label{table-velocities}
\end{table}

\begin{deluxetable}{lrrrrrr}
\tablecaption{Emission Lines}
\tablewidth{0pc}
\tablenum{3}
\tablehead{
\colhead{Name}&
\colhead{PC1}&
\colhead{PC2}&
\colhead{F1}&
\colhead{F2}&
\colhead{PC3}&
\colhead{PC4}
}
\startdata
He D3      & 5866.6 & 5872.1 &  5874.7 & 5876.7 & 5877.6 & 5879.1 \\
He6678     & 6675.2 & 6676.1 &  6677.8 & 6678.9 & 6679.1 & 6680.2 \\
H$\alpha$  & 6557.7 & 6560.0 &  6561.0 & 6564.6 & 6566.2 & 6568.4 \\
H$\beta$   & 4854.7 & 4857.5 &  4860.3 & 4863.3 & 4865.0 & 4870.0 \\
H$\gamma$  & 4335.3 & 4336.9 &  4339.4 & 4341.8 & 4343.4 & 4347.4 \\
H$\delta$  & 4097.2 & 4100.3 &  4100.9 & 4102.5 & 4102.9 & 4105.0 \\
H$\epsilon$& 3969.2 & 3969.6 &  3969.8 & 3970.9 & 3971.2 & 3972.2 \\
H6         & 3887.0 & 3887.8 &  3888.5 & 3889.5 & 3890.0 & 3891.7 \\
\enddata
\end{deluxetable}

\begin{table}
\dummytable\label{table-defs}
\end{table}

\begin{deluxetable}{rrrrrrrrrrrr}
\tablewidth{0pc}
\tablenum{4}
\tablecaption{Equivalent Widths}
\tablehead{
\colhead{NN} &
\colhead{H$\alpha$ } &
\colhead{HeD3} &
\colhead{He6678 } &
\colhead{H$\beta$} &
\colhead{H$\gamma$} &
\colhead{H$\delta$} &
\colhead{H$\epsilon$} &
\colhead{H6} &
\colhead{N$_{{\rm H}\alpha}$} & 
\colhead{$\sigma_{{\rm H}\alpha}$} & 
\colhead{$\sigma_{{\rm H}\beta}$} 
}
\startdata
 7 & -0.396 & 0.034 & -0.057 & -0.130 & \nodata & \nodata & \nodata & \nodata &  1 &  \nodata &  \nodata \\ 
 11 & -0.570 & 0.054 & -0.149 & -0.184 & \nodata & \nodata & \nodata & \nodata &  1 &  \nodata &  \nodata \\ 
 12 & -0.585 & 0.045 & -0.156 & -0.208 & \nodata & \nodata & \nodata & \nodata &  1 &  \nodata &  \nodata \\ 
 19 & 0.108 & -0.287 & 0.258 & 1.701 & \nodata & \nodata & \nodata & \nodata &  3 &  0.190 &  3.681 \\ 
 38 & -0.325 & 0.075 & -0.065 & -0.053 & \nodata & \nodata & \nodata & \nodata &  1 &  \nodata &  \nodata \\ 
 41 & -0.224 & -0.020 & 0.078 & 1.513 & \nodata & \nodata & \nodata & \nodata &  6 &  0.056 &  3.511 \\ 
 43 & -0.326 & -0.011 & 0.066 & -0.223 & \nodata & \nodata & \nodata & \nodata &  1 &  \nodata &  \nodata \\ 
 54 & -0.569 & 0.033 & -0.187 & -0.231 & \nodata & \nodata & \nodata & \nodata &  2 &  0.090 &  0.037 \\
\enddata
\end{deluxetable}

\begin{table}
\dummytable\label{table-widths}
\end{table}

\begin{deluxetable}{rlr}
\tablecaption{Rapid Rotators}
\tablewidth{0pt}
\tablenum{5}
\tablehead{
\colhead{NN} & 
\colhead{Name} &
\colhead{$v \sin i$ (km/s)} 
}
\startdata
381 & G 173-039 & 30 \\
556 & GJ 1054A & 30 \\
802 & G 039-029 & 30 \\
1279 & LHS 5134 & 35 \\
1759 & Steph 932 & 30 \\
2128 & G 165-008 & 80 \\
3453 & G 188-038 & 37 \\
3198 & Gl 791.2 & 35 \\
\enddata
\end{deluxetable}

\begin{table}
\dummytable\label{table-rotation}
\end{table}

\pagestyle{empty}
\begin{deluxetable}{rll}
\tablecaption{Double-Lined Binaries}
\tablewidth{0pt}
\tablenum{6}
\tablehead{
\colhead{NN} & 
\colhead{Name}  &
\colhead{Reference}
}
\startdata
 129 &  Gl 29.1  & Strassmeier et al. (1993) \\ 
 556 &  GJ 1054A & \nodata        \\ 
 558 &  LTT 1480 & \nodata      \\ 
 656 &  Gl 148   &Tokovinin (1991)      \\ 
 1138 &  Gl 268  &  Strassmeier et al. (1993)  \\ 
 1196 &  Gl 278C & Strassmeier et al. (1993)   \\ 
 1328 &  GJ 2069A & Delfosse et al. (1999)  \\ 
 1405 &  G 041-014 & Delfosse et al. (1999)  \\ 
 1759 & Steph 932 & \nodata \\
 1890 &  Gl 455  &\nodata  \\ 
 2017 &  Gl 487  & Delfosse et al. (1999)  \\ 
 2128 &  G 165-008 & \nodata \\
 2242 &  LTT 5580 & \nodata \\
 2316 &  LHS 2952 & \nodata  \\ 
 2616 &  Gl 630.1A & Strassmeier et al. (1993) \\ 
 2935 &  GJ 1230AC &  Delfosse et al. (1999)  \\
 2976 &  Gl 735 & Strassmeier et al. (1993)  \\ 
 3278 &  Gl 815A & Strassmeier et al. (1993)   \\ 
 3561 &  Gl 866A & Delfosse et al. (1999)  \\ 
 3562 &  Gl 867A & Strassmeier et al. (1993)   \\ 
 3701 &  GJ 1284 & \nodata \\ 
 3785 &  G 273-185 & \nodata   \\ 
\nodata & G 173-018 & \nodata \\
\enddata
\end{deluxetable}

\begin{table}
\dummytable\label{table-sb2}
\end{table}

\pagestyle{empty}
\begin{deluxetable}{rlrl}
\tablecaption{Candidate Single-Lined Binaries: Multiple observations}
\tablewidth{0pt}
\tablenum{7}
\tablehead{
\colhead{NN} & 
\colhead{Name}  &
\colhead{$\sigma$} &
\colhead{Reference}
}
\startdata
 853 & LP 476-207 & 4.0 & Delfosse et al. (1999) \\
 1554 & Gl 372 & 24.0 & Harlow (1996) \\
 2097 & Gl 508 A & 5.5 & \nodata \\
 2708 & G 203-47 & 34.2 & Delfosse et al. (1999) \\
 3563 & Gl 867 B & 20.9 & \nodata \\
\enddata
\end{deluxetable}

\begin{table}
\dummytable\label{table-sb1}
\end{table}

\pagestyle{empty}
\begin{deluxetable}{rlcl}
\tablecaption{Candidate Single-lined Binaries: Velocity differences}
\tablewidth{0pt}
\tablenum{8}
\tablehead{
\colhead{NN} & 
\colhead{Name} &
\colhead{$\Delta v$ (km/s)} &
\colhead{Source}  
}
\startdata
 1019 & Gl 228 A & -54 & Upgren and Caruso (1988) \\
 1361 & Gl 319 A & -8 &  Upgren and Harlow (1996)\\ 
 1938 & Gl 462 & -9 & Upgren and Caruso (1988) \\
 2157 & Gl 521.1 & -25 & Stauffer \& Hartmann (1986) \\
 2235 & Gl 540 & -10 &  Stauffer \& Hartmann (1986)\\
 2486 & Gl 600 & 10 &  Upgren and Caruso (1988) \\
\enddata
\end{deluxetable}

\begin{table}
\dummytable\label{table-sb1diff}
\end{table}

\pagestyle{empty}
\begin{deluxetable}{rrrrcc}
\tablecaption{Candidate wide binaries}
\tablewidth{0pt}
\tablenum{9}
\tablehead{
\colhead{Star 1} & 
\colhead{V$_{rad}$} &
\colhead{Star 2} & 
\colhead{V$_{rad}$} &
\colhead {Separation } &
\colhead{$\Delta V / \sigma$} \\
\colhead{} & \colhead{kms$^{-1}$} &
\colhead{} & \colhead{kms$^{-1}$} &
\colhead {A.U.}& \colhead {} }
\startdata
Gl 48 & 0.5 & G 22 C & -4.9 & 225400 & 3.1\\
 & & Gl 22 B & -3.4 &  \\
Gl 49 & -5.0 & Gl 51 & -7.3 & 3530 & 1.5 \\
Gl 140 A & 24.0 & Gl 140 C & 19.1 & 2310 &3.3 \\
G 192-11 & 2.0 & G 192-12 & 2.8 & 2965 & 0.5 \\
Gl 469 & -10.2 & Gl 471 & 15.1 & 52170 & 17 \\
Wo 9490 A & -29.2 & Wo 9490 C & -26.6 & 3538 & 1.7 \\
Gl 720 A & -30.4 & Gl 720 B & -32.2 & 2312 &1.7 \\
Gl 897 A & -0.8 & Gl 898 & 0.7 & 6068 & 1.0 \\
\enddata
\end{deluxetable}

\begin{table}
\dummytable\label{table-wide}
\end{table}

 \pagestyle{empty}
\begin{deluxetable}{rcrrrclrl}
\tablecaption{M and L/T Dwarf Systems}
\tablewidth{0pt}
\tablenum{10}
\tablehead{
\colhead{NN} & 
\colhead{M Dwarf Name}  &
\colhead{Sp.}  &
\colhead{V-I}  &
\colhead{H$\alpha$ (\AA)}  &
\colhead{L/T Dwarf Name}  &
\colhead{Sp. Type}  &
\colhead{Age(Gyr)}  &
\colhead{Ref}
}
\startdata
1021 & Gl 229A& M0.5 & 2.01 & -0.5 & Gl 229B & T & $>0.32$ & N95 \\
2368 & Gl 570BC & M1.0 & 2.12 & -0.4 & Gl 570D & T & $>0.40$ & B00 \\
\nodata & G 196-3 & M2.5 & 2.34 & 4 &  G196-3B & L & $<0.64$ & R98 \\
3 & GJ 1001A & M3.5 & 2.69 & $<1$ & GJ 1001B & L & $>1.4$ & G99 \\ 
2358 & Gl 569 A & M2 & 2.24 & 1.29 & Gl 569Ba/Bb & M8/L0? & $<0.52$ & M00 \\
\enddata
\tablecomments{References: N95: Nakajima et al. (1995); R98: Rebolo et al. 
(1998); B00: Burgasser et al. (2000); G99: Goldman et al. (1999);
M00: Mart{\'{\i}}n et al. (2000) }
\end{deluxetable}

\begin{table}
\dummytable\label{table-bd}
\end{table}

%\clearpage

%--------------------------BIBLIOGRAPHY---------------------------

\clearpage

\end{document}